\RequirePackage{etoolbox}
\csdef{input@path}{%
 {sty/}
 {img/}
}%
\csgdef{bibdir}{bib/}

\documentclass[ba]{imsart}
\pubyear{0000}
\volume{00}
\issue{0}
\doi{0000}
\firstpage{1}
\lastpage{1}

\usepackage{amsthm}
\usepackage{amsmath}
\usepackage{natbib}
\usepackage{bbm}
\usepackage[colorlinks,citecolor=blue,urlcolor=blue,filecolor=blue,backref=page]{hyperref}
\usepackage{graphicx}
\usepackage{multicol}
\usepackage{float}

\startlocaldefs
\endlocaldefs

\begin{document}


\begin{frontmatter}
\title{Dirichlet Process Mixture Models for Modeling and
        Generating Synthetic Versions of Nested Categorical Data}

\runtitle{}
\begin{aug}
\author{\fnms{Jingchen} \snm{Hu}\thanksref{addr1,t1,m1}\ead[label=e1]{jihu@vassar.edu}},
\author{\fnms{Jerome P.} \snm{Reiter}\thanksref{addr2,t1,t2,m1,m2}\ead[label=e2]{jerry@stat.duke.edu}}
\and
\author{\fnms{Quanli} \snm{Wang}\thanksref{addr2,t2,m2}
\ead[label=e3]{quanli@stat.duke.edu}
\ead[label=u1,url]{http://www.foo.com}}

\runauthor{J. Hu et al.}

\address[addr1]{Department of Mathematics and Statistics, Vassar College, Box 27, Poughkeepsie, NY 12604
    \printead{e1} 
}

\address[addr2]{Department of Statistical Science, Duke
    University, Durham, NC 27708-0251
    \printead{e2}
}

\address[addr3]{Department of Statistical Science, Duke
    University, Durham, NC 27708-0251
    \printead{e3}
}

\thankstext{t1}{National Science Foundation CNS-10-12141}
\thankstext{t2}{National Science Foundation SES-11-31897}
\thankstext{t3}{Arthur P. Sloan Foundation G-2015-2-166003}

\end{aug}

\begin{abstract}

We present a Bayesian model for estimating the joint distribution of multivariate
categorical data when units are nested within groups.  Such data arise
frequently in social science settings, for example, people living in households.  The model assumes that (i) each group is a member of a
group-level latent class, and (ii) each unit is a member of a unit-level latent class
 nested within its group-level latent class.  This structure allows
 the model to capture dependence among units in the same group.  It
 also facilitates simultaneous modeling of variables at both group and
 unit levels.  We develop a version of the model that assigns zero probability to groups and
units with physically impossible combinations of variables.
We apply the model to estimate multivariate
 relationships in a subset of  the American Community
 Survey. Using  the estimated model, we generate synthetic
 household data that could be disseminated as redacted public use
 files.
Supplementary materials for this article are available online.

\end{abstract}

\begin{keyword}
\kwd{Confidentiality}
\kwd{Disclosure}
\kwd{Latent}
\kwd{Multinomial}
\kwd{Synthetic}
\end{keyword}

\end{frontmatter}






\section{Introduction}\label{introduction}
In many settings, the data comprise units nested within groups (e.g.,
people within households), and include categorical variables measured at the unit level (e.g.,
individuals' demographic characteristics) and at the group level
(e.g., whether the family owns or rents their home). A typical analysis
goal is to estimate multivariate relationships among the categorical
variables, accounting for the hierarchical structure in the data.

To estimate joint distributions with multivariate categorical data,
many analysts rely on mixtures of products of multinomial distributions, also known as
latent class models. These models
assume that each unit is a member of an unobserved cluster, and that variables follow independent multinomial
distributions within clusters.  Latent class
models can be estimated via  maximum likelihood
\citep{goodman1974biometrika} and Bayesian approaches
\citep{ishwaran:james, jainneal, dunsonxing}. Of particular note, \citet{dunsonxing} present a
nonparametric Bayesian version of the latent class model, using a
Dirichlet process mixture (DPM) for the prior distribution.  The DPM
prior distribution is appealing, in that (i) it has full support on
the space of joint distributions for unordered categorical variables,
ensuring that the model does not restrict dependence structures {\em a priori}, and
(ii) it fully incorporates uncertainty about the effective
number of latent classes in posterior inferences.

For data  nested within groups, however, standard latent class models
may not offer accurate estimates of joint distributions.
In particular, it may not be appropriate to treat the units in the same group as independent;
for example, demographic variables like age, race, and sex of individuals in the same household are
clearly dependent.  Similarly, some
combinations of units may be physically impossible to place in the
same group, such as a daughter who is older than her biological father.  Additionally, every unit in a group must
have the same values of group-level variables, so that one cannot
simply add multinomial kernels for the group-level variables.

In this article, we present a  Bayesian mixture model for nested categorical data.  The
model assumes that (i) each group is a member of a group-level latent class, and
 (ii) each unit is a member of a unit-level latent class
 nested within its group-level latent class.  This structure encourages the model to
 cluster groups into data-driven types, for example, households with
 children where everyone has
the same race. This in turn allows for dependence among units in the same group.
The nested structure also facilitates simultaneous modeling of
variables at both group and unit levels.  We refer to the model as the
nested data Dirichlet process mixture of
products of multinomial distributions (NDPMPM).
We present two versions of the
NDPMPM: one that gives support to all  configurations of
groups and units, and one that assigns zero probability to groups and
units with physically impossible combinations of variables (also known
as structural zeros in the categorical data analysis literature).

The NDPMPM is similar to the latent class models proposed by
\citet{vermunt03, vermunt08}, who also uses two layers of
latent classes to model nested categorical data.  These models use a fixed number
of classes as determined by a model selection criterion (e.g., AIC or
BIC), whereas the NDPMPM allows uncertainty in the effective number of
classes at each level. The NDPMPM also is similar to the latent
class models in \citet{benninketal} for nested data, especially to what they call the
``indirect model.'' The indirect model regresses a
single group-level outcome on group-level and individual-level predictors, whereas the NDPMPM is used for estimation
of the joint distribution of multiple group-level and individual-level
variables.  To the best of our knowledge, the models of \citet{vermunt03,
  vermunt08} and \citet{benninketal}
do not account for groups with physically
impossible combinations of units.

One of our primary motivations in developing the NDPMPM is to develop a method for generating
redacted public use files for household data, specifically for the
variables on the United States decennial census. Public use files in which confidential data values are
replaced with draws from  predictive distributions are known in the
disclosure limitation literature as synthetic datasets
\citep{rubin:1993, little:1993, raghu:rubin:2001, reiter:2002a,
  reiter:raghu:07}.  Synthetic data techniques  have been used to
create several high-profile public use data products,
including the Survey of Income and Program Participation
\citep{abowd06}, the Longitudinal Business Database \citep{lbdisr},
the American Community Survey group quarters data \citep{hawalaacs},
and the OnTheMap application \citep{onthemap}. None of these products
involve synthetic household data. In these products, the synthesis
strategies are based on chains of generalized linear models for independent
individuals, e.g., simulate variable $x_1$ from some parametric
model $f(x_1)$, $x_2$ from some parametric model $f(x_2|x_1)$, etc.
We are not aware of any synthesis models appropriate for nested categorical data
like the decennial census variables.

As part of generating the synthetic data, we evaluate disclosure risks
using the measures suggested in \citet{hureiterwang14}. Specifically,
we quantify the posterior probabilities that intruders can learn
values from the confidential data given the released synthetic data,
under assumptions about the intruders' knowledge and attack strategy.
This is the only strategy we know of for evaluating statistical
disclosure risks for nested categorical data.  To save space, the
methodology and results for the disclosure
risk evaluations are presented in the supplementary material only.
To summarize very briefly, the analyses suggest that synthetic data
generated from the NDPMPM have low disclosure risks.

The remainder of this article is organized as follows. In Section \ref{model}, we
present the NDPMPM model when all configurations of groups
and units are feasible. In Section
\ref{augmented}, we present a data augmentation strategy for
estimating a  version of the NDPMPM that puts zero probability on
impossible combinations.
In Section \ref{sims}, we illustrate and evaluate the NDPMPM models using household
demographic data from the American Community Survey (ACS).
In particular, we use posterior predictive distributions from the
NDPMPM models to generate synthetic datasets, and compare results of representative
analyses done with the synthetic and original data.  In Section \ref{conclusion}, we conclude with discussion of implementation of the
proposed models.

\section{The NDPMPM Model}\label{model}

As a working example, we suppose the data include $N$ individuals
residing in only one of $n < N$ households, where $n$ (but not $N$) is
fixed by design.
For $i = 1, \dots, n$, let $n_i\geq 1$ equal the number of individuals in house $i$, so that $\sum_{i=1}^n n_i = N$.
For $k = 1, \dots, p$,  let $X_{ijk} \in \{1, \dots, d_k\}$ be the value of categorical variable $k$ for person $j$ in household $i$, where $i = 1, \dots, n$ and $j = 1, \dots, n_i$.
For $k=p+1, \dots, p+q$, let $X_{ik} \in \{1, \dots, d_k\}$ be the
value of categorical variable $k$ for household $i$, which is assumed
to be identical for all $n_i$ individuals in household $i$.  We let
one of the variables in $X_{ik}$ correspond to the household size
$n_i$; thus, $N$ is a random variable. For now, we assume no impossible combinations of
variables within individuals or households.

We assume that each household belongs to some group-level latent
class, which we label with $G_i$, where $i=1, \dots, n$. Let
$\pi_g = \mbox{Pr}(G_i = g)$ for any class $g$; that is, $\pi_g$ is the probability that household $i$
belongs to class $g$ for every household.
For any $k \in \{p+1, \dots, p+q\}$ and any value $c \in \{1,\dots,
d_{k}\}$, let $\lambda_{gc}^{(k)} = \mbox{Pr}(X_{ik}=c \mid
G_{i}=g)$ for any class $g$; here, $\lambda_{gc}^{(k)}$ is the same
value for every household in class $g$.
 For computational expediency, we truncate the number of group-level
latent classes at some sufficiently large value $F$.  Let
$\mathbf{\pi} = \{\pi_1, \dots, \pi_F\}$, and let $\mathbf{\lambda} = \{\lambda_{gc}^{(k)}: c=1,\dots,d_{k}; k=p+1,\dots, p+q;
g=1,\dots, F\}$.

Within each household class, we assume that each individual member belongs
to some individual-level latent class, which we label with
$M_{ij}$, where $i=1, \dots, n$ and $j=1, \dots, n_i$.  Let
$\omega_{gm} = \mbox{Pr}(M_{ij} = m \mid G_i = g)$ for any class $(g,
m)$; that is, $\omega_{gm}$ is the conditional probability
that individual $j$ in household $i$ belongs to individual-level class
$m$ nested within group-level class $g$, for every individual.
For any $k \in \{1, \dots, p\}$ and any value $c \in \{1,\dots, d_{k}\}$, let $\phi_{gmc}^{(k)} =
\mbox{Pr}(X_{ijk}=c \mid (G_i, M_{ij}) = (g,m))$;
here, $\phi_{gmc}^{(k)}$ is the same value for every individual in class $(g,m)$.
Again for computational expediency, we truncate the number of
individual-level latent classes within each $g$ at some sufficiently
large number $S$ that is common across all $g$. Thus, the truncation results in a total of $F\times S$
latent classes used in computation.
Let  $\mathbf{\omega} = \{\omega_{gm}: g=1, \dots, F; m=1, \dots, S\}$, and let
$\mathbf{\phi} = \{\phi_{gmc}^{(k)}: c=1,\dots,d_{k}; k=1,\dots,p;
g=1,\dots, F; m =1,\dots,S\}$.

We let both the $q$ household-level variables and $p$ individual-level
variables follow independent, class-specific multinomial
distributions.
Thus, the model for the data and corresponding latent classes in the NDPMPM is
\begin{eqnarray}
	\label{lca-xh} X_{ik} \mid G_i, \mathbf{\lambda} &{\sim}& \textrm{Multinomial}(\lambda_{G_i1}^{(k)}, \dots,
		\lambda_{G_i d_{k}}^{(k)})  \nonumber \\ &\,\,\,\,&\textrm{for all
                } i,\, k = p+1, \dots, p+q\\
\label{lca-xi} X_{ijk} \mid G_i, M_{ij}, n_i, \mathbf{\phi} &{\sim}& \textrm{Multinomial}(\phi_{G_iM_{ij}1}^{(k)}, \dots,
		\phi_{G_iM_{ij}d_k}^{(k)}) \nonumber \\ &\,\,\,\,& \textrm{for all }
                i, j, k=1, \dots, p\\
	\label{lca-eta} G_i \mid \mathbf{\pi} &\sim& \textrm{Multinomial}(\pi_1, \dots, \pi_F) \,\,\,\, \textrm{for all } i,\\	
	\label{lca-psi} M_{ij} \mid G_i, n_i, \mathbf{\omega} &\sim& \textrm{Multinomial}(\omega_{G_i1},\dots,\omega_{G_iS}) \,\,\,\, \textrm{for all } i,j,
\end{eqnarray}
where each multinomial distribution has sample size equal to one and number of levels implied by the dimension
of the corresponding probability vector.  We allow the
multinomial probabilities for individual-level classes to differ by
household-level class. One could impose additional structure on the
probabilities, for example, force them to be equal across classes as
suggested in \citet{vermunt03, vermunt08}; we do not pursue such
generalizations here.

We condition on $n_{i}$ in \eqref{lca-xi} and \eqref{lca-psi} so that the entire model can
be interpreted as a generative model for households; that is, the size of the household could be sampled from
\eqref{lca-xh}, and once the size is known the characteristics of the
household's individuals could be sampled from \eqref{lca-xi}.
The distributions in \eqref{lca-xi} and \eqref{lca-psi} do not depend on
$n_i$ other than to fix the number of people in the household; that
is, within any $G_i$, the distributions of all parameters do not
depend on $n_i$. This encourages   borrowing strength across
households of different sizes while simplifying computations.

As prior distributions on $\pi$ and $\omega$, we use the truncated
stick breaking  representation of the Dirichlet process
\citep{seth94}. We have
\begin{align}
\label{prior1}	\pi_g &= u_{g}\prod_{f<g}(1-u_{f}) \,\,\,\, \textrm{for } g=1,
        \dots, F\\ 	
	u_{g}& \sim \textrm{Beta} (1,\alpha) \,\,\,\, \textrm{for } g=1, \dots, F-1, \,\,\,\, u_F=1\\
\alpha &\sim \textrm{Gamma}(a_{\alpha}, b_{\alpha})\\
    \omega_{gm} &= v_{gm}\prod_{s<m}(1-v_{gs}) \,\,\,\, \textrm{for } m=1, \dots, S\\
	v_{gm}  &\sim \textrm{Beta} (1,\beta_g) \,\,\,\, \textrm{for } m=1, \dots, S-1, \,\,\,\, v_{gS}=1\\	
    \beta_g  &\sim \textrm{Gamma}(a_{\beta}, b_{\beta}). \label{betag}
\end{align}
The prior distribution in \eqref{prior1}--\eqref{betag} is similar to the
truncated version of the nested Dirichlet process prior distribution of
\citet{RodriguezDunsonGelfand2008} based on conditionally conjugate
prior distributions (see Section 5.1 in their article).
The prior distribution in \eqref{prior1}--\eqref{betag} also shares
characteristics with the enriched Dirichlet process prior distribution
of \citet{WadeMongPetrone11}, in that (i) it gets around the
limitations caused by using a single precision parameter $\alpha$ for
the mixture probabilities, and (ii) it allows different mixture
components for different variables.

As prior distributions on $\lambda$ and $\phi$, we use independent
Dirichlet distributions,
\begin{align}
	\mathbf{\lambda}_{g}^{(k)}=(\lambda_{g1}^{(k)},\dots,\lambda_{gd_{k}}^{(k)})&\sim \textrm{Dir} (a_{k1}, \dots,a_{kd_{k}})\\
    \mathbf{\phi}_{gm}^{(k)}=(\phi_{gm1}^{(k)},\dots,\phi_{gmd_k}^{(k)})&\sim \textrm{Dir} (a_{k1}, \dots,a_{kd_{k}}).
\end{align}
One can use data-dependent prior distributions for setting
each $(a_{k1}, \dots, a_{k_{d_k}})$, for example, set it equal to the empirical marginal
frequency.  Alternatively, one can set $a_{k1} = \dots = a_{kd_{k}} = 1$ for all $k$ to correspond to
uniform distributions.  We examined both approaches and found no
practical differences between them for our applications; see the
supplementary material. In the applications, we present
results based on the empirical marginal frequencies.  Following
\citet{dunsonxing} and \citet{si:reiter:13}, we set $(a_{\alpha} =.25,  b_{\alpha}=.25)$ and $(a_{\beta} =.25,  b_{\beta}=.25)$,
which represents a small prior sample size and hence vague specification for the Gamma distributions.
We estimate the posterior distribution of
all parameters using a blocked Gibbs sampler \citep{ishwaran:james,
  si:reiter:13}; see the supplement for the relevant full conditionals.

Intuitively, the NDPMPM seeks to cluster households with
similar compositions. Within the pool of individuals in any household-level
class, the model seeks to cluster individuals with similar
characteristics.  Because individual-level latent class assignments
are conditional on household-level latent class assignments, the model
induces dependence among individuals in the same
household (more accurately, among
  individuals in the same household-level cluster). To see this mathematically, consider the
expression for the joint distribution for variable $k$ for two
individuals $j$ and $j'$ in the same household $i$.  For any $(c, c')
\in \{1, \dots, d_k\}$, we have
\begin{equation}
\label{jointprob}
Pr(X_{ijk}=c, X_{ij'k}=c') =
\sum_{g=1}^F \left(\sum_{m=1}^S \phi_{gmc}^{(k)} \omega_{gm}
\sum_{m=1}^S \phi_{gmc'}^{(k)} \omega_{gm}\right) \pi_g.
\end{equation}
Since $Pr(X_{ijk}=c) = \sum_{g=1}^F \sum_{m=1}^S
  \phi_{gmc}^{(k)} \omega_{gm}\pi_g$  for any $c \in \{1, \dots,
  d_k\}$, the $Pr(X_{ijk}=c, X_{ij'k}=c') \neq Pr(X_{ijk}=c)
  Pr(X_{ij'k}=c')$.

Ideally we fit enough latent classes to capture key features in the data while
keeping computations as expedient as possible.  As a strategy for doing so, we have found it convenient
to start an MCMC chain with reasonably-sized values of $F$ and $S$, say $F = S = 10$.  After
convergence of the MCMC chain, we check how many latent classes at the household-level
and individual-level are occupied across the MCMC iterations. When the
numbers of occupied household-level classes hits $F$, we
increase $F$.  When this is not the case but the number of occupied
individual-level classes hits $S$, we try increasing $F$
alone, as the increased number of household-level latent classes may
sufficiently capture heterogeneity across households as to make $S$ adequate.
When increasing $F$ does not help, for example there are too many different
types of individuals, we increase $S$, possibly in addition
to $F$.  We emphasize that these types of titrations are useful primarily to reduce computation time;
analysts always can set $S$ and $F$ both to be very large
so that they are highly likely to exceed the number of occupied classes in initial runs.

It is computationally convenient to set $\beta_g = \beta$ for all $g$ in \eqref{betag}, as
doing so reduces the number of parameters in the model. Allowing
$\beta_g$ to be class-specific offers additional flexibility, as the
prior distribution of the household-level class probabilities can vary
by class. In our evaluations of the model on the ACS data, results
were similar whether we used a common or distinct values of $\beta_g$.

\section{Adapting the NDPMPM for Impossible Combinations}\label{augmented}

The models in Section \ref{model} make no restrictions on the
compositions of groups or individuals. In many contexts this is
unrealistic. Using our working example, suppose that the data include
a variable that characterizes relationships among individuals
in the household, as the ACS does.  Levels of this variable include  household
head, spouse of household head, parent of the household head, etc.  By
definition, each household  must contain exactly one household head.
Additionally, by definition (in the ACS), each household
head must be at least 15 years old. Thus, we require a version of the NDPMPM that enforces zero probability for
any household that has zero or multiple household heads, and any household headed by
someone younger than 15 years.

We need to modify the likelihoods in \eqref{lca-xh} and \eqref{lca-xi} to enforce zero probability for impossible combinations.
Equivalently, we need to truncate the support of the NDPMPM.  To
express this mathematically, let $\mathcal{C}_h$ represent all
combinations of individuals and households of size $h$, including
impossible combinations; that is, $\mathcal{C}_h$ is the Cartesian product $\Pi_{k=p+1}^{p+q}
(1, \dots, d_k) \left(\Pi_{j=1}^{h} \Pi_{k=1}^p (1, \dots,
  d_k)\right)$.  For any household with $h$ individuals, let
$\mathcal{S}_h \subset \mathcal{C}_h$ be the set of combinations that should have
zero probability, i.e., $Pr(X_{ip+1}, \dots, X_{ip+q}, X_{i11},
\dots, X_{ihp} \in \mathcal{S}_h) =0$. Let $\mathcal{C} = \bigcup_{h \in
  \mathcal{H}}{\mathcal{C}_h}$ and $\mathcal{S} = \bigcup_{h \in
  \mathcal{H}}{\mathcal{S}_h}$, where  $\mathcal{H}$ is the set of all household
sizes in the observed data.  We define a random variable for all the
data for person $j$ in household $i$ as
${\mathbf{X}^*_{ij}}=(X^*_{ij1}, \dots, X^*_{ijp}, X^*_{ip+1}, \dots,
X^*_{ip+q})$, and a random variable for all data in household $i$ as
$\mathbf{X}^*_{i} = (\mathbf{X}^*_{i1}, \dots, \mathbf{X}^*_{in_i})$.
Here, we write a superscript $*$ to indicate that the random variables have support only on $\mathcal{C} - \mathcal{S}$; in contrast,
we use $\mathbf{X}_{ij}$ and $\mathbf{X}_{i}$ to indicate the corresponding random variables
with unrestricted support on $\mathcal{C}$.
Letting  $\mathcal{X}^*$ be the sampled data from $n$ households, i.e., a realization of
$(\mathbf{X}^*_1, \dots, \mathbf{X}^*_n)$,
the likelihood component of the
truncated NDPMPM model, $p({\mathcal{X}^*}|\theta)$,  can be written
 as proportional to $L(\mathcal{X}^* \mid \theta)=$
\begin{equation}
\label{NDPMPMtruncated}
\prod_{i=1}^n \sum_{h\in\mathcal{H}}\left(\mathbbm{1}\{n_i=h\}\mathbbm{1}\{{\mathbf{X}^*_i}\notin {\mathcal{S}_h}\}\sum_{g=1}^F \left(\prod_{k=p+1}^{p+q}\lambda_{gX^*_{ik}}^{(k)}\left(\prod_{j=1}^{h}\sum_{m=1}^S\prod_{k=1}^p\phi_{gmX^*_{ijk}}^{(k)} \omega_{gm}\right)\right) \pi_g\right)
\end{equation}
where $\theta$ includes all parameters of the model described in
Section \ref{model}. Here,
$\mathbbm{1}\{.\}$ equals one when the condition inside the $\{\}$ is true and equals zero otherwise.

For all $h \in \mathcal{H}$, let $n_{*h} = \sum_{i=1}^n \mathbbm{1}\{n_i=h\}$ be
the number of households of size $h$ in $\mathcal{X}^*$.  Let $\pi_{0h}(\theta)=Pr({\mathbf{X}_{i}}\in
{\mathcal{S}_h}|\theta)$, where  $\mathbf{X}_{i}$ is the random
variable with unrestricted support.  The normalizing constant in the likelihood
in \eqref{NDPMPMtruncated} is $\prod_{h\in\mathcal{H}}(1-\pi_{0h}(\theta))^{n_{*h}}$.
Hence,  we seek to compute the posterior distribution
\begin{equation}
\label{NDPMPMtruncatedpost}
p(\theta|\mathcal{X}^*, T(\mathcal{S})) \propto p(\mathcal{X}^* \mid \theta)p(\theta) = \frac{1}{\prod_{h\in\mathcal{H}}(1-\pi_{0h}(\theta))^{n_{*h}}} L(\mathcal{X}^* \mid \theta) p(\theta).
\end{equation}
The
$T(\mathcal{S})$ emphasizes that the density is for the truncated NDPMPM, not the density from Section \ref{model}.

The Gibbs sampling strategy from Section \ref{model} requires conditional independence across individuals and variables,
and hence unfortunately is not appropriate as a means to estimate the posterior distribution.
Instead, we follow the general approach of \citet{manriquereiterjcgs}. The basic idea
is to treat the observed data $\mathcal{X}^*$, which we assume includes only feasible
households and individuals (e.g., there are no reporting
errors that create impossible combinations in the observed data), as a sample from
an augmented dataset $\mathcal{X}$
of unknown size.  We assume $\mathcal{X}$
arises from an NDPMPM model that
does not restrict the characteristics of households or individuals; that is, all combinations of households
and individuals are allowable in the augmented sample.  With this
conceptualization, we can construct a Gibbs sampler that
appropriately assigns zero probability to combinations in $\mathcal{S}$ and results in draws of $\theta$ from \eqref{NDPMPMtruncatedpost}. Given a draw of $\theta$,
we draw $\mathcal{X}$ using a negative binomial sampling scheme.  For each
stratum $h \in \mathcal{H}$ defined by unique household sizes
in $\mathcal{X}^*$, we repeatedly simulate households with individuals from the
 untruncated NDPMPM model, stopping when the number of simulated feasible households matches $n_{*h}$.
We make $\mathcal{X}$ comprise $\mathcal{X}^*$ and the generated households that fall in $\mathcal{S}$.
Given a draw of $\mathcal{X}$, we draw $\theta$ from the
NDPMPM model as in Section \ref{model}, treating $\mathcal{X}$ as if it were collected data.
The full conditionals for this
sampler, as well as a proof that it generates draws from \eqref{NDPMPMtruncatedpost}, are provided in the supplement.

\section{Using the NDPMPM to Generate Synthetic Household Data}\label{sims}

We now illustrate the ability of the NDPMPM to estimate joint distributions for subsets
of household level and individual level variables. Section \ref{simnozero} presents results for
a scenario where the variables are free of structural zeros (i.e., $\mathcal{S}=\emptyset$), and Section
\ref{simzero} presents results for a scenario with impossible
combinations.

We use subsets of variables selected from the public use files for the
ACS. As brief background, the purpose of the ACS is to enable estimation of
population demographics and housing characteristics for the entire
United States. The questionnaire is sent to about 1 in 38 households.
It includes questions about the individuals living in the household
 (e.g., their ages, races, incomes) and about the characteristics of
 the housing unit (e.g., number of bedrooms,
 presence of running water or not, presence of a telephone line or
 not). We use only data from non-vacant households.

In both simulation scenarios, we treat data from the public use files
as populations, so as to have known population values, and take
simple random samples from them on which we estimate the NDPMPM models. We
use the estimated posterior predictive distributions to create
simulated versions of the data, and compare analyses of the simulated
data to the corresponding analyses based on the observed data and the
constructed population values.

If we act like the samples from the constructed populations are
confidential and cannot be shared as is, the simulated datasets can be
viewed as redacted public use file, i.e., synthetic data.  We generate $L$ synthetic datasets,
$\textbf{Z}=(\textbf{Z}^{(1)}, \dots, \textbf{Z}^{(L)})$, by sampling $L$ datasets from
the posterior predictive distribution of a NDPMPM model.
We generate synthetic data so that the number of households of any size $h$ in each $\textbf{Z}^{(l)}$
exactly matches $n_{*h}$.  This improves the quality of the synthetic
data by ensuring that the total number of individuals and household
size distributions match in $\textbf{Z}$ and $\mathcal{X}^*$.  As a result, $\textbf{Z}$
comprises partially synthetic data \citep{little:1993, reiterpartsyn}, even though every released $Z_{ijk}$ is a
simulated value.

To make inferences with $\textbf{Z}$ we use the approach in \citet{reiterpartsyn}.
Suppose that we seek to estimate some scalar quantity $Q$.  For $l=1, \dots,
L$, let $q^{(l)}$ and $u^{(l)}$ be respectively the point estimate of $Q$
and its associated variance estimate computed with $\textbf{Z}^{(l)}$.  Let
$\bar{q}_L = \sum_l q^{(l)}/L$; $\bar{u}_L = \sum_l u^{(l)}/L$;
$b_L = \sum_l (q^{(l)} - \bar{q}_L)^2/(L-1)$; and $T_L = \bar{u}_L + b_L/L$. We make
inferences about $Q$ using the $t-$distribution, $(\bar{q}_L - Q) \sim
t_{v}(0, T_L)$, with $v = (L-1)(1+L\bar{u}_L/ b_L)^2$ degrees of
freedom.


\begin{table}[t]
\centering
\begin{tabular}{ll}
Description & Categories \\ \hline
Ownership of dwelling& 1 = owned or being bought, 2 = rented\\
House acreage &1 = house on less than 10 acres,\\
                &2 = house on 10 acres or more\\
Household income &1 = less than 25K, 2 = between 25K and 45K, \\
                &3 = between 45K and 75K,\\
                & 4 = between 75K and 100K, 5 = more than 100K\\
Household size & 1 = 1 person, 2 = 2 people, etc.\\
Age &1 = 18, 2 = 19, \dots, 78 = 95\\
Gender & 1 = male, 2 = female \\
Recoded general race code & 1 = white alone, 2 = black alone, \\
                            &3 = American Indian/Alaska Native alone, \\
                            &4 = Asian or Pacific Islander alone,  \\
                            &5 = other, 6 = two or more races \\
Speaks English &1 = does not speak English, 2 = speaks English\\
Hispanic origin &1 = not Hispanic, 2 = Hispanic\\
Health insurance coverage&1 = no, 2 = yes\\
Educational attainment &1 = less than high school diploma, \\
                        &2 = high school diploma/GED/alternative credential, \\
                        &3 = some college, 4 = bachelor's degree, \\
                        &5 = beyond bachelor's degree\\
Employment status&1 = employed, 2 = unemployed, 3 = not in labor force\\
Migration status, 1 year&1 = in the same house, 2 = moved within state, \\
                        &3 = moved between states, 4 = abroad one year ago\\
Marital status&1 = married spouse present, \\
                &2 = married spouse absent, 3 = separated, \\
                &4 = divorced, 5 = widowed, \\
                &6 = never married/single\\ \hline
\end{tabular}
\caption{Subset of variables in the empirical illustration without  structural
  zeros.  The first four variables are household-level
  variables, and the last ten variables are individual-level
  variables. \label{varsnozeros}}
\end{table}

\subsection{Illustration without structural zeros}\label{simnozero}

For this scenario, we use data from the 2012 ACS public use file
\citep{ACS:20112012} to construct a population with 308769
households.
From this we take a simple random sample of $n=10000$ households. We use
the four household-level variables and ten individual-level
variables summarized in Table \ref{varsnozeros}. We select these variables purposefully
to avoid structural zeros.  Household sizes range from one to nine, with
$(n_{*1}, \dots, n_{*9}) = (2528, 5421, 1375, 478, 123, 52, 16, 5,
2)$.  This sample of $n$ households includes $N=20504$ individuals.  We treat
income and age as unordered categorical variables; we discuss adapting
the model for ordered categorical variables in Section \ref{conclusion}.

We run the MCMC sampler for the NDPMPM model of Section \ref{model}
for 10000 iterations, treating the first 5000
iterations as burn-in.  We set $(F, S) = (30, 10)$ and use a common
$\beta$.   The posterior mean of the number of occupied
household-level classes is 27 and ranges from 25 to 29.
Within household-level classes, the posterior number of occupied individual-level classes ranges from 5 to 8.
To monitor convergence of the MCMC sampler, we focus of $\pi$, $\alpha$, and $\beta$. As a check on the choice of $(F, S)$, we also estimated the model with $(F, S) = (50, 50)$.
We found similar results for both the number of occupied classes and the posterior predictive distributions; see the supplement for details.

\begin{figure}[t]
\centering
\includegraphics[scale=0.35]{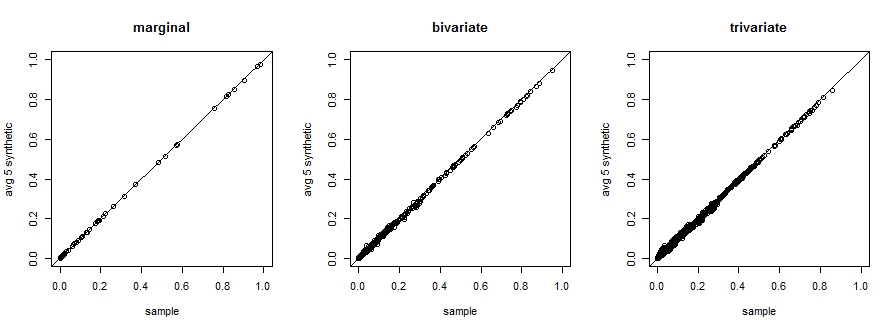}
\caption{Marginal, bivariate and trivariate probabilities computed in the sample and synthetic
  datasets for the illustration without structural zeros.
 Restricted to categories with expected counts equal to at least 10. Point
  estimates from both sets of data are similar, suggesting that the NDPMPM fits the data well.}
\label{app1}
\end{figure}

We generate $\textbf{Z}^{(l)}$ by
sampling a draw of $(\mathbf{G}, \mathbf{M}, \lambda, \phi)$ from the posterior distribution.
For each household $i=1, \dots, n$, we generate its synthetic
household-level attributes, $(X_{i p+1}^{(l)}, \dots, X_{i p+q}^{(l)})$, from \eqref{lca-xh} using $G_i$ and
the corresponding probabilities in $\mathbf{\lambda}$.
For each individual $j=1, \dots, n_i$ in each household, we generate the synthetic individual-level attributes,
$(X_{ij1}^{(l)}, \dots, X_{ijp}^{(l)})$, from \eqref{lca-xi} using $M_{ij}$ and the corresponding
probabilities in $\mathbf{\phi}$. We repeat this process
$L=5$ times, using approximately independent draws of
parameters obtained from iterations that are far apart in the MCMC chain.

To evaluate the quality of the NDPMPM model, we compare the
relationships among the variables in the original and synthetic
datasets to each other, as is typical in synthetic data evaluations,
as well as to the corresponding population values. We consider the marginal distributions of all variables,
bivariate distributions of all possible pairs of variables, and
trivariate distributions of all possible triplets of variables.  We restrict the plot to
categories where the expected count in samples of 10000 households is at least 10. Plots in Figure \ref{app1}
display each $\bar{q}_5$ plotted against its corresponding empirical
probability in the original data for all parameters.  As evident in the figures, the synthetic point estimates
are close to those from the original data, suggesting that
the NDPMPM accurately estimates the relationships among the variables.
Both sets of point estimates are close to the
corresponding probabilities in the population, as we show in the supplement.

\begin{table}[t]
\centering
\begin{tabular}{lrrrr}
  &Q&Original &NDPMPM&DPMPM\\ \hline
  All same race && &\\
$\,\,\,\,\,$ $n_i = 2$&.928 &(.923, .933)& (.847, .868)&(.648, .676)\\
$\,\,\,\,\,$  $n_i = 3$&.906 &(.889, .901)& (.803, .845)&(.349, .407)\\
$\,\,\,\,\,$ $n_i = 4$&.885&(.896, .908)&(.730, .817)&(.183, .277)\\
  All white, rent&.123 & (.115, .128)&(.110, .126)&(.052, .062)\\
  All white w/ health insur.&.632 &(.622, .641)&(.582, .603)&(.502, .523)\\
  All married, working&.185 & (.177, .192)&(.171, .188)&(.153, .168)\\
  All w/ college degree&.091 &(.086, .097)& (.071, .082)&(.067, .077)\\
  All w/ health coverage&.807 &(.800, .815)& (.764, .782)&(.760, .777)\\
  All speak English&.974 &(.969, .976)&(.959, .967)&(.963, .970)\\
  Two workers in home &.291 &(.282, .300) & (.289, .309)&(.287, .308)\\  \hline
\end{tabular}
\caption{95\% confidence intervals in the original and synthetic data
  for selected probabilities that depend on within household
  relationships. Results for illustration without structural zeros.
  Intervals for probability that all family members are the same race
  are presented only for households of size two, three, and four
because of inadequate sample sizes for $n_i>4$.
The
quantity $Q$ is the value in the constructed population of 308769
households.}
\label{app1HHraceandtype}
\end{table}

We also examine several probabilities that depend on values for individuals in the same
household, that is, they are affected by within-household
relationships. As evident in Table \ref{app1HHraceandtype}, and not
surprisingly given the sample size, the point estimates from the original
sampled data are close to the values in the constructed population. For most quantities the
synthetic data point and interval estimates  are similar to those based on the original sample,
suggesting that the  NDPMPM model
has captured the complicated within household structure reasonably
well.  One exception is the percentage of households with everyone
of the same race: the NDPMPM underestimates these
percentages.  Accuracy worsens as household size increases.
This is partly explained by sample sizes, as
$n_{*3} = 1375$ and $n_{*4}=478$, compared to $n_{*2} = 5421$.
We also ran a simulation with $n= 50000$ households comprising
$N=101888$ individuals sampled
randomly from the same constructed population, in which $(n_{*1}, \dots, n_{*10}) =
(12804, 27309, 6515, 2414, 630, 229, 63, 26, 8, 2)$.
For households with $n_i=3$, the 95\% intervals from the synthetic and
original data are, respectively, (.870, .887) and (.901, .906); for
households of size $n_i=4$, the 95\% intervals from the synthetic and
original data are, respectively, (.826, .858) and (.889, .895). Results
for the remaining probabilities in Table \ref{app1HHraceandtype} are
also improved.

As a comparison, we also generated synthetic datasets using a non-nested DPMPM model
\citep{dunsonxing} that ignores the household clustering.  Not
surprisingly, the DPMPM results in substantially less accuracy for
many of the probabilities in Table \ref{app1HHraceandtype}. For example, for
the percentage of households of size $n_i=4$ in which all members have
the same race, the DPMPM results in a 95\% confidence interval of
(.183, .277), which is quite unlike the (.896, .908) interval
in the original data and far from the population value of .885. The DPMPM also struggles for other quantities
involving racial compositions.  Unlike the NDPMPM model, the DPMPM
model treats each observation as independent, thereby ignoring the
dependency among individuals in the same household.
We note that we obtain similar results with nine other independent
samples of 10000 households, indicating that the differences between
the NDPMPM and DPMPM results in Table \ref{app1HHraceandtype} are not
reflective of chance error.

\subsection{Illustration with structural zeros}\label{simzero}

\begin{table}[t]
\centering
\begin{tabular}{ll}
Description & Categories \\ \hline
Ownership of dwelling& 1 = owned or being bought (loan), 2 = rented\\
Household size & 2 = 2 people, 3 = 3 people, 4 = 4 people\\
Gender & 1 = male, 2 = female \\
Race &1 = white, 2 = black, \\
    &3 = American Indian or Alaska Native, \\
    &4 = Chinese, 5 = Japanese, \\
    &6 = other Asian/Pacific Islander, \\
    &7 = other race,  8 = two major races, \\
    &9 = three/more major races \\
Hispanic origin (recoded)&1 = not Hispanic, 2 = Mexican, \\
                        &3 = Puerto Rican, 4 = Cuban, 5 = other\\
Age (recoded)&1 = 0 (less then one year old), 2 = 1, \dots, \\
                &94 = 93\\
Relationship to the household head&1 = head/householder, 2 = spouse, 3 = child, \\
                                    &4 = child-in-law, 5 = parent, 6 = parent-in-\\
                                    &law, 7 = sibling, 8 = sibling-in-law, \\
                                    &9 = grandchild, 10 = other relatives, \\
                                    &11 = partner, friend, visitor, \\
                                    &12 = other non-relatives\\ \hline
\end{tabular}
\caption{Subset of variables used in the illustration with structural
  zeros.  The first two variables are household-level
  variables, and the last five variables are individual-level variables.\label{varszeros}}
\end{table}

For this scenario, we use data from the 2011 ACS public use file
\citep{ACS:20112012} to construct the population. We select variables
to mimic those on the U.\ S.\ decennial census, per the motivation
described in Section \ref{introduction}.  These include a variable
that explicitly indicates relationships among individuals within the
same household.  This variable creates numerous and complex patterns of impossible combinations.
For example, each household can have only one head who must
be at least 16 years old, and biological children/grandchildren must be younger than their
parents/grandparents.
We use the two household-level variables and five individual-level
variables  summarized in Table \ref{varszeros}, which  match those on the decennial census questionnaire.
We exclude households with only one individual because these individuals by definition must be
classified as household heads, so that we have no need to model the
family relationship variable.  To generate synthetic data
for households of size $n_i=1$,  one could use non-nested versions of
latent class models \citep{dunsonxing, manriquereiterjcgs}.   We also
exclude households with $n_i > 4$ for presentational and computational
convenience.

The constructed population comprises 127685 households, from which we
take a simple random sample of $n=10000$ households.  Household sizes are  $(n_2, n_3, n_4)=(5370, 2504, 2126)$.
The $10000$ households comprise $N=26756$ individuals.

We fit the the truncated NDPMPM model of Section \ref{augmented},
using all the variables in Table \ref{varszeros} as $X_{ijk}$ or $X_{ik}$ in the model.
We run the MCMC sampler for 10000 iterations, treating the first 6000
iterations as burn-in. We set $(F, S) = (40, 15)$ and
use a common $\beta$. The posterior mean of the number of
household-level classes occupied by households in $\mathcal{X}^*$ is 28 and ranges from 23 to 36.  Within household-level classes, the posterior number of individual-level classes
occupied by individuals in $\mathcal{X}^*$ ranges from 5 to 10.
To check for convergence of the MCMC chain, we look at trace plots of
$\mathbf{\pi}$, $\alpha$,  $\beta$, and $n_0$.  The plots for
$(\mathbf{\pi}, \alpha, \beta)$ suggest good mixing; however, the plot for
$n_0$ exhibits non-trivial auto-correlations.
Values of $n_0$ are around
$8.0 \times 10^5$ near the 6000th and 10000th iterations of the chain,
with a minimum around $7.2 \times 10^5$ near the 6500th iteration and a maximum around $9.3 \times 10^5$ near the 9400th iteration.
As a byproduct of the MCMC sampler, at each MCMC iteration we create
$n$ households that satisfy all constraints.  We use
these households to form each $\textbf{Z}^{(l)}$, where $l=1, \dots, 5$,
selecting from five randomly sampled, sufficiently separated
iterations.

\begin{figure}[t]
\centering
\includegraphics[scale=0.35]{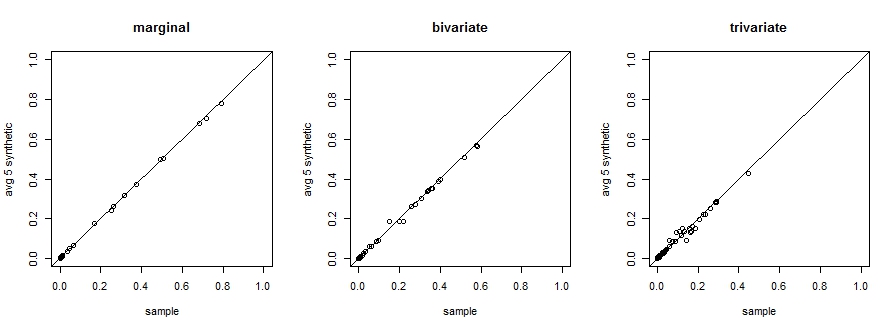}
\caption{Marginal, bivariate and trivariate distributions probabilities computed in the sample and synthetic
  datasets in illustration with structural zeros. Restricted to categories with expected counts equal to at least 10. Point
  estimates from both sets of data are similar, suggesting that
  that the truncated NDPMPM fits the data reasonably well.}
\label{app2}
\end{figure}

As in Section \ref{simnozero}, we evaluate the marginal distributions of all variables,
bivariate distributions of all possible pairs of variables, and
trivariate distributions of all possible triplets of variables, restricting to categories where the expected counts are at least 10.
Plots in Figure \ref{app2} display each $\bar{q}_5$ plotted against its corresponding
estimate from the original data, the latter of which are close to
the population values (see the supplementary material).  The point estimates are quite
similar, indicating that the NDPMPM captures relationships among the variables.

\begin{table}[t]
\centering
\begin{tabular}{lrrrrr}
  &Q& Original& NDPMPM & NDPMPM &NDPMPM\\
  &&&truncate&untruncate&rej samp\\ \hline
All same race & && \\
$\,\,\,\,\,$ $n_i=2$&.906 &(.900, .911)&(.858, .877)&(.824, .845)&(.811, .840)\\
$\,\,\,\,\,$ $n_i=3$&.869 &(.871, .884)&(.776, .811)&(.701, .744)&(.682, .723)\\
$\,\,\,\,\,$ $n_i=4$&.866 &(.863, .876)&(.756, .800)&(.622, .667)&(.614, .667)\\
   Spouse present&.667 &(.668, .686)&(.630, .658)&(.438, .459)&(.398, .422)\\
   Spouse w/ white HH&.520&(.520, .540)&(.484, .510)&(.339, .359)&(.330, .356)\\
   Spouse w/ black HH&.029 &(.024, .031)&(.022, .029)&(.023, .030)&(.018, .025)\\
   White cpl&.489 &(.489, .509)&(.458, .483)&(.261, .279)&(.306, .333)\\
   White cpl, own&.404 &(.401, .421)&(.370, .392)&(.209, .228)&(.240, .266)\\
   Same race cpl&.604 &(.603, .622)&(.556, .582)&(.290, .309)&(.337, .361)\\
   White-nonwhite cpl&.053&(.049, .057)&(.048, .058)&(.031, .039)&(.039, .048)\\
   Nonwhite cpl, own&.085&(.079, .090)&(.068, .079)&(.025, .033)&(.024, .031)\\
   Only mother&.143 &(.128, .142)&(.103, .119)&(.113, .126)&(.201, .219)\\
   Only one parent&.186&(.172, .187)&(.208, .228)&(.230, .247)&(.412, .435)\\
   Children present&.481&(.473, .492)&(.471, .492)&(.472, .492)&(.566, .587)\\
   Parents present&.033&(.029, .036)&(.038, .046)&(.035, .043)&(.011, .016)\\
   Siblings present&.029&(.022, .028)&(.032, .041)&(.027, .034)&(.029, .039)\\
   Grandchild present&.035&(.028, .035)&(.032, .041)&(.035, .043)&(.024, .031)\\
   Three generations&.043&(.036, .043)&(.042, .051)&(.051, .060)&(.028, .035)\\
$\,\,\,\,\,\,$ present & & & & & \\ \hline
\end{tabular}
\caption{95\% confidence intervals in the original and synthetic data
  for selected probabilities that depend on within household
  relationships. Results for illustration with structural zeros.
``NDPMPM truncate" uses the model from Section \ref{augmented}. ``NDPMPM untruncate" uses the model from Section \ref{model}. ``NDPMPM rej samp" uses the model from Section \ref{model} but rejecting any proposed synthetic observation that fails to respect the structural zeros.
``HH'' means household head, and ``cpl'' means couple.  The quantity
$Q$ is the value in the full constructed population of 127685  households.}
\label{app2summary}
\end{table}

Table \ref{app2summary} compares original and
synthetic 95\% confidence intervals for selected probabilities involving
within-household relationships.  We choose a wide range of household
types involving multiple household level and individual level
variables. We include quantities that depend explicitly on the
``relationship to household head'' variable, as these should be
particularly informative about how well the truncated NDPMPM model
estimates probabilities directly impacted by structural zeros. As
evident in Table \ref{app2summary}, estimates from the original sample data
are generally close to the corresponding population values. Most
intervals from the synthetic data are similar to those from the
original data, indicating that the truncated NDPMPM model
captures within-household dependence structures reasonably
well. As in the simulation with no structural zeros, the truncated
NDPMPM model has more difficulty
capturing dependencies for the larger households, due to
smaller sample sizes and more complicated within-household
relationships.

For comparison, we also generate synthetic data using the NDPMPM model from Section \ref{model}, which does not
account for the structural zeros.  In the column labeled
``NDPMPM untruncate'', we use the NDPMPM model and
completely ignore structural zeros, allowing the synthetic data to
include households with impossible combinations.  In the column
labeled ``NDPMPM rej samp'', we ignore structural zeros when
estimating model parameters but use rejection sampling at the data
synthesis stage to ensure that no simulated households include
physically impossible combinations.   As seen in Table
\ref{app2summary}, the interval estimates from the truncated NDPMPM
generally are more accurate than those based on the other two
approaches.
When structural zeros most directly impact the probability, i.e., when the ``relationship to household head'' variable is involved, the performances of ``NDPMPM untruncate'' and ``NDPMPM rej samp'' are substantially degraded.

\section{Discussion}\label{conclusion}

The MCMC sampler for the NDPMPM in Section \ref{model} is computationally expedient.
However, the MCMC sampler for
the truncated NDPMPM in Section \ref{augmented} is computationally intensive.
The primary bottlenecks in the computation arise from simulation of $\mathcal{X}$.  When the probability mass in the region defined by
$\mathcal{S}$ is large compared to the probability mass in the region
defined by $\mathcal{C} - \mathcal{S}$, the MCMC can sample
many households with impossible combinations before getting $n$ feasible
ones.  Additionally, it can be time consuming to check whether or not a
generated record satisfies all constraints in $\mathcal{S}$.  These bottlenecks can be
especially troublesome when $n_i$ is large for many households.  To
reduce running times, one can parallelize many steps in the sampler
(which we did not do). As examples,
the generation of augmented records and the checking of constraints
can be spread over many processors.  One also can reduce computation
time by putting an upper bounds on the size of $\mathcal{X}$ (that is still much larger
than $n$).  Although this results in an approximation to the Gibbs
sampler, this still could yield  reasonable inferences or synthetic datasets,
particularly when many records in $\mathcal{X}$ end up in clusters
with few data points from $\mathcal{X}^*$.

Conceptually, the methodology can be readily extended to handle other
types of variables. For example, one could replace the multinomial
kernels with continuous kernels (e.g., Gaussian distributions) to
handle numerical variables.  For ordered categorical variables, one
could use a probit specification \cite{albertchib} or the rank likelihood \cite[][Ch. 12]{Bayesian:Hoff}.
For mixed data, one could use the Bayesian joint model for multivariate continuous and categorical
variables developed in \citet{murray:reiter:mixed}.  Evaluating the
properties of such models is a topic for future research.

We did not take advantage of prior information when estimating the
models.  Such information might be known, for example, from other data
sources. Incorporating prior information in latent class models is
tricky, because we need to do so in a way that does not distort
conditional distributions.  \citet{schifeling:reiter:BA} presented a
simple approach to doing so for non-nested latent class models, in which
the analyst appends to the original data partially complete, pseudo-observations with empirical
frequencies that match the desired prior distribution.  If one had
prior information on household size jointly with some other variable,
say individuals' races, one could follow the approach of
\citet{schifeling:reiter:BA} and augment the collected data with
partially complete households.  When the prior information does not
include household size, e.g., just a marginal distribution of race, it
is not obvious how to incorporate the prior information in a
principled way.


Like most joint models, the NDPMPM generally is not appropriate for
estimating multivariate distributions with data from complex sampling designs.
This is because the model reflects the distributions in the
observed data, which might be collected by differentially sampling
certain subpopulations.  When design variables are categorical and
are available for the entire population (not just the sample),
analysts can use the NDPMPM as an engine for Bayesian finite
population inference \citep[][Ch. 8]{gcrs:3rd}.  In this case, the analyst includes the design
variables in the NDPMPM, uses the implied, estimated conditional
distribution to impute many copies of the non-sampled records' unknown survey values
given the design variables, and computes quantities of interest on
each completed population.  These completed-population quantities
summarize the posterior distribution. Absent this information, there is
no consensus on the ``best'' way to incorporate survey weights in
Bayesian joint mixture models.  \citet{dunson:kuni:weights}
present a computationally convenient approach that
uses only the survey weights for sampled cases. A similar approach could be applied for nested categorical data.
Evaluating this approach, as well as other adaptations of ideas proposed in the literature, is a worthy topic for future research.

The truncated NDPMPM also assumes the observed data do not include errors that create theoretically impossible combinations of values.
When such faulty values are present, analysts should edit and impute corrected values, for example, using the
\citet{fellegi:holt} paradigm popular with statistical agencies.  Alternatively, one could add a stochastic measurement error model to the truncated NDPMPM,  as done by \citet{kimetal} for continuous data and \citet{manrique:reiter:2016} for non-nested categorical data.  While conceptually feasible, this is not a trivial extension.  The NDPMPM is already
computationally intensive; searching over the huge space of possible error localizations
could increase the computational burden substantially. This
suggests one would need alternatives to standard MCMC algorithms for model fitting.

\newpage
Supplementary materials
\section{Introduction}

Section 2 describes the full conditionals for the Gibbs sampler for both NDPMPM models. Section 3
presents a proof that the sampler for the NDPMPM with structural zeros gives draws from the
posterior distribution of $\theta$ under the truncated model.
Sections 4 to 7 present the results of the assessments of disclosure risks for the synthetic data illustrations in
Section 4 of the main text. We describe the methodology for assessing risks in Section 4 and
the computational methods in Section 5. We summarize the disclosure risk evaluations for the scenario without and with structural zeros
in Section 6 and Section 7, respectively.
Section 8 presents plots of point estimates versus the population values for both the synthetic and the original sample data, as described in Section 4 of the main text.
Section 9 presents and compares results using an empirical prior and uniform prior distribution for the multinomial parameters in the no structural zeros simulation.
Section 10 presents and compares results using $(F, S) = (50, 50)$ and $(F, S) = (30, 10)$ in the no structural zeros simulation.

\section{Full conditional distributions for MCMC samplers}
We present the full conditional distributions used in the Gibbs
samplers for the versions of the NDPMPM with and without structural
zeros. In both presentations, we assume common $\beta$ for all
household-level clusters.


\subsection{NDPMPM without structural zeros}
\begin{align*}
\intertext{- Sample $G_i \in \{1,\dots,F\}$ from a multinomial distribution with sample size one and probabilities}
Pr(G_i&=g|-) = \frac{\pi_g\{\prod_{k=p+1}^{q}\lambda_{gX_{ik}}^{(k)}(\prod_{j=1}^{n_i}\sum_{m=1}^S\omega_{gm}\prod_{k=1}^{p}\phi_{gmX_{ijk}}^{(k)})\}}{\sum_{f=1}^{F}\pi_f\{\prod_{k=p+1}^{q}\lambda_{fX_{ik}}^{(k)}(\prod_{j=1}^{n_i}\sum_{m=1}^{S}\omega_{fm}\prod_{k=1}^{p}\phi_{fmX_{ijk}}^{(k)})\}}.
\intertext{- Sample $M_{ij} \in \{1,\dots,S\}$ given $G_i$ from a multinomial distribution with sample size one and probabilities}
Pr(M_{ij}&=m|-) =
\frac{\omega_{G_im}\prod_{k=1}^{p}\phi_{G_imX_{ijk}}^{(k)}}{\sum_{s=1}^{S}\omega_{G_is}\prod_{k=1}^{p}\phi_{G_isX_{ijk}}^{(k)}}.
\intertext{- Set $u_{F}=1$. Sample $u_g$ from the Beta distribution for $g=1,\dots,F-1$, where}
(u_g|-) &\sim \textrm{Beta}(1+\sum_{i=1}^{n}\mathbbm{1}(G_i=g),\alpha+\sum_{f=g+1}^{F}\sum_{i=1}^{n}\mathbbm{1}(G_i=f))\\
\pi_g &= u_g \prod_{f<g}(1-u_f).
\intertext{- Set $v_{gM}=1$. Sample $v_{gm}$ from the Beta distribution for $m=1,\dots,S-1$, where}
(v_{gm}|-) &\sim \textrm{Beta}(1+\sum_{i=1}^{n}\mathbbm{1}(M_{ij}=m,G_i=g),\beta+\sum_{s=m+1}^{S}\sum_{i=1}^{n}\mathbbm{1}(M_{ij}=s,G_i=g))\\
\omega_{gm} &= v_{gm} \prod_{s<m}(1-v_{gs}).
\intertext{- Sample $\lambda_g^{(k)}$ from the Dirichlet distribution for $g=1,\dots,F$, and $k=p+1,\dots,q$, where}
(\lambda_g^{(k)}|-) &\sim
\textrm{Dir}(a_{k1}+\sum_{i|G_i=g}^{n}\mathbbm{1}(X_{ik}=1),\dots,a_{kd_{k}}+\sum_{i|G_i=g}^{n}\mathbbm{1}(X_{ik}=d_{k})).\\
\intertext{- Sample $\phi_{gm}^{(k)}$ from the Dirichlet distribution for $g=1,\dots,F$, $m=1,\dots,S$ and $k=1,\dots,p$, where}
(\phi_{gm}^{(k)}|-) &\sim \textrm{Dir}(a_{k1}+\sum_{i,j|\substack{G_i=g,\\M_{ij}=m}}^{n, n_i}\mathbbm{1}(X_{ijk}=1),\dots,
a_{kd_k}+\sum_{i,j|\substack{G_i=g,\\M_{ij}=m}}^{n, n_i}\mathbbm{1}(X_{ijk}=d_k)).\\
\intertext{- Sample $\alpha$ from the Gamma distribution,}
(\alpha|-) &\sim \textrm{Gamma}(a_{\alpha}+F-1,b_{\alpha}-\sum_{g=1}^{F-1}log(1-u_g)).\\
\intertext{- Sample $\beta$ from the Gamma distribution,}
(\beta|-) &\sim \textrm{Gamma}(a_{\beta}+F*(S-1), b_{\beta}-\sum_{m=1}^{S-1}\sum_{g=1}^Flog(1-v_{gm})).\\
\end{align*}

\subsection{NDPMPM with structural zeros}\label{samplertrunc}

Let $\mathcal{X}^0$ include observations that are not admissible (they fail structural zero constraints).
Let $\mathbf{G}^0$ and $\mathbf{M}^0$ be the latent class membership indicators for these records.
Let the total number of households of size $h$ in $\mathcal{X}$ be written as $(n_{*h} + n_{0h})$,
where $n_{0h}$ is the number households of size $h$ generated in $\mathcal{X}^0$.

In each MCMC iteration, we have to sample $(n_{0h}, \{(\mathcal{X}^0_i, G_i^0, M_{ij}^0) : i = 1, \dots, n_{0h}, j = 1, \dots, h\})$
for each $h \in \mathcal{H}$.
We do so by means of a rejection
sampler.  To begin, we initialize $\mathcal{X}^0 =\mathbf{G}^0 =\mathbf{M}^0 = \emptyset$ at each
MCMC iteration.  For each $h \in \mathcal{H}$, we repeat the following
steps.
\begin{itemize}
\item[a.] Set $t_0=0$.  Set $t_1 = 0$.
\item[b.] Sample a value of $G_i$ from a multinomial distribution with sample size one and
$Pr(G_i=g|-) \propto Pr(X_{ik} = h \mid G_i = g) \pi_g$, where $X_{ik}$ corresponds to the variable for household size.
\item[c.] For $j = 1, \dots, h$, sample a value of $M_{ij}$ from a multinomial distribution with sample size one and
$Pr(M_{ij}=m|-) = \omega_{G_{im}}$.
\item[d.] Set $X_{ik} = h$.  Sample remaining household level
  values and all individual level values using (1) and
  (2) from the main text.  Let $\mathcal{X}_{i}^0$ be the simulated value.
\item[e.] If $\mathcal{X}_{i}^0 \in \mathcal{S}_h$, let $t_0 = t_0 + 1$ and
  $\mathcal{X}^0 =\mathcal{X}^0 \cup \mathcal{X}_{i}^0$. Similarly,
let $\mathbf{G}^0 =\mathbf{G}^0 \cup G_{i}$ and $\mathbf{M}^0 =\mathbf{M}^0 \cup \{M_{i1}, \dots, M_{ih}\}$.
 Otherwise
  set $t_1 = t_1 + 1$.
\item[f.] If $t_1 < n_{*h}$, return to Step b.  Otherwise set
  $n_{0h} = t_0$.
\end{itemize}
\begin{align*}
\intertext{- For observations in $\mathcal{X}^*$,
sample $G_i \in \{1,\dots,F\}$ from a multinomial distribution with sample size one and}
Pr(G_i&=g|-) = \frac{\pi_g\{\prod_{k=p+1}^{q}\lambda_{gX_{ik}}^{(k)}(\prod_{j=1}^{n_i}\sum_{m=1}^S\omega_{gm}\prod_{k=1}^{p}\phi_{gmX_{ijk}}^{(k)})\}}{\sum_{f=1}^{F}\pi_f\{\prod_{k=p+1}^{q}\lambda_{fX_{ik}}^{(k)}(\prod_{j=1}^{n_i}\sum_{m=1}^{S}\omega_{fm}\prod_{k=1}^{p}\phi_{fmX_{ijk}}^{(k)})\}}.
\intertext{- For observations in $\mathcal{X}^*$,
sample $M_{ij} \in \{1,\dots,S\}$ given $G_i$ from a multinomial distribution with sample size one and}
Pr(M_{ij}&=m|-) =
\frac{\omega_{G_im}\prod_{k=1}^{p}\phi_{G_imX_{ijk}}^{(k)}}{\sum_{s=1}^{S}\omega_{G_is}\prod_{k=1}^{p}\phi_{G_isX_{ijk}}^{(k)}}.
\intertext{- Set $u_{F}=1$. Let $n_0 = \sum_h n_{0h}$. Sample $u_g$ from the Beta distribution for $g=1,\dots,F-1$, where}
(u_g|-) &\sim \textrm{Beta}(1+\sum_{i=1}^{n}\mathbbm{1}(G_i=g),\alpha+\sum_{f=g+1}^{F}\sum_{i=1}^{n+n_0}\mathbbm{1}(G_i=f))\\
\pi_g &= u_g \prod_{f<g}(1-u_f).
\intertext{- Set $v_{gM}=1$. Sample $v_{gm}$ from the Beta distribution for $m=1,\dots,S-1$, where}
(v_{gm}|-) &\sim \textrm{Beta}(1+\sum_{i=1}^{n+n_0}\mathbbm{1}(M_{ij}=m,G_i=g),\beta+\sum_{s=m+1}^{S}\sum_{i=1}^{n+n_0}\mathbbm{1}(M_{ij}=s,G_i=g))\\
\omega_{gm} &= v_{gm} \prod_{s<m}(1-v_{gs}).
\intertext{- Sample $\lambda_g^{(k)}$ from the Dirichlet distribution for $g=1,\dots,F$, and $k=p+1,\dots,q$, where}
(\lambda_g^{(k)}|-) &\sim
\textrm{Dir}(a_{k1}+\sum_{i|G_i=g}^{n+n_0}\mathbbm{1}(X_{ik}=1),\dots,a_{kd_{k}}+\sum_{i|G_i=g}^{n+n_0}\mathbbm{1}(X_{ik}=d_{k})).\\
\intertext{- Sample $\phi_{gm}^{(k)}$ from the Dirichlet distribution for $g=1,\dots,F$, $m=1,\dots,S$ and $k=1,\dots,p$, where}
(\phi_{gm}^{(k)}|-) &\sim \textrm{Dir}(a_{k1}+\sum_{i,j|\substack{G_i=g,\\M_{ij}=m}}^{n+n_0, n_i}\mathbbm{1}(X_{ijk}=1),\dots,a_{kd_k}+\sum_{i,j|\substack{G_i=g,\\M_{ij}=m}}^{n+n_0, n_i}\mathbbm{1}(X_{ijk}=d_k)).\\
\intertext{- Sample $\alpha$ from the Gamma distribution,}
(\alpha|-) &\sim \textrm{Gamma}(a_{\alpha}+F-1,b_{\alpha}-\sum_{g=1}^{F-1}log(1-u_g)).\\
\intertext{- Sample $\beta$ from the Gamma distribution,}
(\beta|-) &\sim \textrm{Gamma}(a_{\beta}+F*(S-1), b_{\beta}-\sum_{m=1}^{S-1}\sum_{g=1}^Flog(1-v_{gm})).\\
\end{align*}

\section{Proof that sampler converges to correct distribution in the truncated model}

In this section, we state and prove a result that ensures draws of $\theta$ from the sampler in Section 3 of the main text correspond to draws from
the posterior distribution, $p(\theta | \mathcal{X}^*, T(\mathcal{S}))$. The proof follows the strategy
in \citet{manriquereiterjcgs}. A key difference is that our MCMC algorithm proceeds separately for each $h$,
generating households from the untruncated model until reaching $n_{*h}$ feasible households.

We begin by introducing notation for the augmented data, $\mathcal{X}$. Recall that
$\mathcal{X}$ is a draw from a NDPMPM model without restrictions, i.e., all
combinations of household and individual variables are allowed.
We write $\mathcal{X} = (\mathcal{X}^1,\mathcal{X}^0)$, where $\mathcal{X}^1$ includes
observations that are admissible (no structural zeros) and $\mathcal{X}^0$ includes observations that are
not admissible (they fail structural zero constraints).

Each record in $\mathcal{X}$ is associated with a household-level and individual-level
 latent class assignment.  
Let ${\mathbf{G}}^1 = (G^1_1, \dots, G^1_n)$ and
${\mathbf{M}}^1=\{(M^1_{i1},\dots,M^1_{in_i}), i=1,\dots,n\}$ include all the
latent class assignments corresponding to households and individuals in $\mathcal{X}^1$.
Let $\mathbf{G}^0=(G^0_1, \dots, G^0_{n_0})$ and
${\mathbf{M}}^0=\{(M^0_{i1}, \dots, M^0_{in_i}), i=1,\dots,n_0\}$ include all the  latent class assignments corresponding to
the $n_0$ cases in $\mathcal{X}^0$.


We seek to prove that one can obtain samples from
$p(\theta|{\mathcal{X}}^*, T(\mathcal{S}))$ in the truncated NDPMPM from a sampler for
$f(\theta,\mathbf{G}^1,\mathbf{G}^0,\mathbf{M}^1,\mathbf{M}^0,\mathcal{X}^0,
\{n_{0h}: h \in  \mathcal{H}\} \mid \mathcal{X}^1)$ under an
untruncated NDPMPM model.  
Put formally, we want to prove the following theorem
\vspace{12pt}

\noindent {\bf Theorem 1:}
{\em Let $\mathcal{X}^*$ comprise $n$ randomly sampled households from the truncated NDPMPM in (15) of the main text.  Let $\mathcal{X}^1$
be generated from the NDPMPM without any concern over structural zeros, i.e., the model from Section 2 of the main text, so
that no element of $\mathcal{X}^1 \in \mathcal{S}$.  Assume that $\mathcal{X}^* = \mathcal{X}^1$.  Let the
prior distribution on each $(n_{*h} + n_{0h})$ be $p(n_{*h} + n_{0h}) \propto 1/(n_{*h} + n_{0h})$.  Then,}
\begin{equation}
\begin{split}
\int
f(\theta,\mathbf{G}^1,\mathbf{G}^0,\mathbf{M}^1,\mathbf{M}^0,\mathcal{X}^0,
\{n_{0h}\} | \mathcal{X}^1)&d{\mathcal{X}}^0 d\mathbf{G}^1 d\mathbf{G}^0 d\mathbf{M}^1
d\mathbf{M}^0 d\{n_{0h}\} \\
&= p(\theta \mid \mathcal{X}^*, T(\mathcal{S})).
\end{split}
\end{equation}
Here, we use integration signs
rather than summation signs to simplify notation.

Before continuing with the proof, we note that the rejection sampling step in the
algorithm for the truncated NDPMPM is equivalent to sampling each $n_{0h}$ from negative binomial distributions.  As
evident in the proof, this distribution arises when one assumes a
specific, improper prior distribution on $(n_{*h}+n_{0h})$ that is independent
of $\theta$, namely $p(n_{*h}+n_{0h}) \propto 1/(n_{*h}+n_{0h})$ for each $h$.  This
improper prior distribution is used solely for computational
convenience, as using other prior distributions would make the full
conditional not negative binomial and hence complicate the sampling of $\mathcal{X}^0$.
 A similar strategy was used by \citet{manriquereiterjcgs}, who
adapted the improper prior suggested by \citet{zaslavsk:meng} and
\citet{zaslavsk:james} for sampling from truncated distributions.

Let $\mathbf{G}_{h} = \{G_{i}: n_i = h\}$ be the household level latent
class assignments of size $h$ households and $\mathbf{M}_{h} =
\{M_{ij}: n_i = h, j=1, \dots, n_i\}$ be the individual level latent class
assignments associated with members of size $h$ households.
We split $\mathbf{G}_{h}$ into $\mathbf{G}_{h}^1$ and
$\mathbf{G}_{h}^0$, representing the values for records in
$\mathcal{X}^1$ and in $\mathcal{X}^0$ respectively. We similarly split $\mathbf{M}_{h}$ into $\mathbf{M}_{h}^1$ and
$\mathbf{M}_{h}^0$.  Let $\mathcal{X}_h^1 = \{\mathcal{X}_i^1 : n_i =
h\}$, and let $\mathcal{X}_h^0 = \{\mathcal{X}_i^0 : n_i =
h\}$.  We emphasize that $\mathcal{X}^1$ is used in all iterations, whereas
$\mathcal{X}^0$ is generated in each iteration of the MCMC sampler.  Using this
notation, we have
\begin{align}
\int
 &f(\theta,\mathbf{G}^1,\mathbf{G}^0,\mathbf{M}^1,\mathbf{M}^0,\mathcal{X}^0,
\{n_{0h}\} | \mathcal{X}^1)d{\mathcal{X}}^0 d\mathbf{G}^1 d\mathbf{G}^0 d\mathbf{M}^1
d\mathbf{M}^0 d\{n_{0h}\} \nonumber \\
 &\propto p(\theta) \prod_{h\in\mathcal{H}} \int
f(\mathcal{X}^1_h, \mathbf{G}^1_h, \mathbf{M}_h^1, \mathcal{X}^0_h, \mathbf{G}^0_h, \mathbf{M}_h^0, n_{0h} \mid \theta) \nonumber \\
&\,\,\,\,\,\,\,\,\,\,\,\,\,\,\,\, d\mathbf{G}_h^1 d\mathbf{M}_h^1 d\mathcal{X}^0_h d\mathbf{G}_h^0 d\mathbf{M}_h^0 dn_{0h}.\label{proofmain}
\end{align}

Extending the generative model in Section 2, we view each $\mathcal{X}^1_h$ as a truncated
sample from the households in $\mathcal{X}$ of size $h$. Let
$\mathcal{A}^1_h$ and $\mathcal{A}^0_h$ be  the set of row indexes of records in $\mathcal{X}_h^1$ and
in $\mathcal{X}_h^0$, respectively. This implies for any given value of
$(n_{*h}+n_{0h})$ that $f(\mathcal{X}^1_h, \mathbf{G}^1_h,
\mathbf{M}_h^1,  \mathcal{X}^0_h, \mathbf{G}^0_h,
\mathbf{M}_h^0 \mid \theta, n_{*h}+n_{0h})$
\begin{align}
 =& \nonumber
{n_{*h}+n_{0h} \choose n_{0h}} \prod_{i \in \mathcal{A}^1_h}
\mathbbm{1}\{{\mathcal{X}}_{i}^1\notin
{{\mathcal{S}}_h}\}f({\mathcal{X}}_{i}^1|
G_{i}^1,\mathbf{M}_{i}^1,\theta)
f(G_{i}^1, \mathbf{M}_{i}^1|\theta)\\
&\prod_{i \in \mathcal{A}^0_h}\mathbbm{1}\{{\mathcal{X}}_{i}^0\in {\mathcal{S}}_h\}
f({\mathcal{X}}_{i}^0 \mid G_{i}^0,\mathbf{M}_{i}^0,\theta) f(G_{i}^0,\mathbf{M}_{i}^0|\theta).\label{mainproof2}
\end{align}

Substituting \eqref{mainproof2} in \eqref{proofmain} and expanding the
integrals, we have
\begin{align*}
& p(\theta) \prod_{h\in\mathcal{H}} \int
f(\mathcal{X}^1_h, \mathbf{G}^1_h, \mathbf{M}_h^1, \mathcal{X}^0_h, \mathbf{G}^0_h, \mathbf{M}_h^0, n_{0h} \mid \theta)
 d\mathbf{G}_h^1 d\mathbf{M}_h^1
 d\mathcal{X}^0_h d\mathbf{G}_h^0, d\mathbf{M}_h^0 dn_{0h}\\
   \begin{split}
   &\propto p(\theta)\prod_{h\in\mathcal{H}}\prod_{i \in \mathcal{A}^1_h} \int \mathbbm{1}\{{\mathcal{X}}_{i}^1\notin {{\mathcal{S}}_h}\}
   f({\mathcal{X}}_{i}^1| G_{i}^1,\mathbf{M}_{i}^1,\theta)f(G_{i}^1|\theta)\prod_{j=1}^{h}f(M_{ij}^1|\theta)dG_i^1 d\mathbf{M}_i^1 \\
   &\qquad \prod_{h\in\mathcal{H}}\sum_{n_{0h}=0}^{\infty} p(n_{*h}+n_{0h}) {n_{*h}+n_{0h} \choose
     n_{0h}}\prod_{i \in \mathcal{A}^0_h}\int \mathbbm{1}\{{\mathcal{X}}_{i}^0\in {\mathcal{S}}_h\}
     f({\mathcal{X}}_{i}^0 \mid G_{i}^0,\mathbf{M}_{i}^0,\theta) \\
     &\qquad f(G_{i}^0,\mathbf{M}_{i}^0|\theta)dG_i^0 d\mathbf{M}_i^0 d\bf{X}_i^0
   \end{split}\\
   \begin{split}
   &= p(\theta)\prod_{h\in\mathcal{H}} \prod_{i \in \mathcal{A}^1_h} \int \mathbbm{1}\{{\mathcal{X}}_{i}^1\notin
   {{\mathcal{S}}_h}\}  f({\mathcal{X}}_{i}^1|
   G_{i}^1,\mathbf{M}_{i}^1,\theta)f(G_{i}^1|\theta)\prod_{j=1}^{h}f(M_{ij}^1|\theta)d\mathbf{G}_i^1
   d\mathbf{M}_i^1 \\
   &\qquad \prod_{h\in\mathcal{H}}\sum_{n_{0h}=0}^{\infty}{n_{*h}+n_{0h}-1 \choose n_{0h}}(\pi_{0h}(\theta))^{n_{0h}}\\
   \end{split}\\
   \begin{split}
   &= p(\theta)\prod_{h\in\mathcal{H}}(\prod_{i \in \mathcal{A}^1_h}\int \mathbbm{1}\{{\mathcal{X}}_{i}^1\notin {{\mathcal{S}}_h}\}f({\mathcal{X}}_{i}^1|G_{i}^1,\mathbf{M}_{i}^1,\theta)f(G_{i}^1|\theta)\prod_{j=1}^{h}f(M_{ij}^1|\theta)d\mathbf{G}_i^1 d\mathbf{M}_i^1\\
   &\qquad (1-\pi_{0h}(\theta))^{-n_{*h}})\\
   \end{split}\\
\end{align*}
From (15) in the main text, this expression is equivalent to
$p(\theta) \prod_{i=1}^n p(\mathcal{X}_{i}^*|\theta)$ when $\mathcal{X}^* = \mathcal{X}^1$, as desired.

Thus, we can obtain samples from the posterior distribution $p(\theta|\mathcal{X}^*, T(\mathcal{S}))$ in
the truncated NDPMPM model from the  sampler for
$f(\theta,\mathbf{G}^*,\mathbf{G}^0,\mathbf{M}^*,\mathbf{M}^0,\mathcal{X}^0,
\{n_{0h}\} \mid \mathcal{X}^*)$
under the unrestricted NDPMPM model.

\section{Disclosure risk measures}

When synthesizing entire household compositions (but keeping household size distributions fixed),
it is nonsensical for intruders to match the proposed synthetic datasets to external files, since there is no unique mapping
of the rows (individuals) in the synthetic datasets $\textbf{Z}$ to the rows in the original data $\textbf{D}$, nor unique
mapping of the households in $\textbf{Z}$ to the households in $\textbf{D}$ (except for household sizes with $n_i = 1$).  We therefore
consider questions of the form: can intruders accurately infer from
$Z$ that some individual or entire household with a
particular set of data values is in the confidential data?  When the
combination of values is unique in the population (or possibly just
the sample), this question essentially asks if intruders can determine
whether or not a specific individual or household is in $D$ \citep{hureiterwang14}.

To describe the disclosure risk evaluations, we follow the presentation of
\citet{hureiterwang14}. We consider two possible attacks on $\textbf{Z}$, namely (i) the intruder seeks to learn whether or not
someone with a particular combination of the $p$
individual-level variables and the $q$ household-level variables is in $\textbf{D}$, and  (ii) an intruder seeks to learn whether or not
an entire household with a particular combination of household-level and individual-level characteristics is in $\textbf{D}$.
For the first scenario, we assume that the intruder knows the values in $\textbf{D}$ for all individuals but the target individual, say
individual $ij$.  We use $\textbf{D}_{-ij}$ to denote the data known to the intruder.
For the second scenario, we assume that the intruder knows the values in $D$ for all households but the target house, say household $i$.
We use $\textbf{D}_{-i}$ to denote the data known to the intruder.
In many cases, assuming the intruder knows $\textbf{D}_{-ij}$ or $\textbf{D}_{-i}$ is conservative; for example, in random
samples from large populations intruders are unlikely to know $N-1$ individuals or $n-1$ households selected in the sample.
 We adopt this strong assumption largely to facilitate computation.  Risks deemed acceptable
under this assumption should be acceptable for weaker intruder knowledge.
We note that assuming the intruder
knows all records but one is related to, but quite distinct from, the
assumptions used in differential privacy \citep{dwork:06}.

Let $T_{ij}$ or $T_i$ be the random variable corresponding to the intruder's guess about the true values of the target.  Let
$t$ generically represent a possible guess at the target, where for simplicity of notation we use a common notation for individual and
household targets.  Let $\mathcal{I}$ represent any information known by the intruder about the
process of generating $Z$, for example meta-data indicating the values
of $F$, $S$ and $(a_{\alpha}, b_{\alpha}, a_{\beta}, b_{\beta})$ for the NDPMPM synthesizer.

For the first type of attack,  we assume the intruder seeks the posterior probability,
\begin{eqnarray}
\rho_{ij}^t = p(T_{ij} = t \mid \textbf{Z}, \textbf{D}_{-ij}, \mathcal{I}) &=& \frac{p(\textbf{Z} \mid T_{ij} = t, \textbf{D}_{-ij}, \mathcal{I})p(T_{ij} = t \mid
\textbf{D}_{-ij}, \mathcal{I})}{\sum_{t
  \in \mathcal{U}}p(\textbf{Z} \mid T_{ij} = t, \textbf{D}_{-ij},
\mathcal{I})p(T_{ij} = t \mid \textbf{D}_{-ij}, \mathcal{I})}\label{fullprob} \\
&\propto& p(\textbf{Z} \mid T_{ij} = t, \textbf{D}_{-ij}, \mathcal{I})p(T_{ij} = t \mid  \textbf{D}_{-ij}, \mathcal{I}) \label{propprob},
\end{eqnarray}
where $\mathcal{U}$ represents the universe of all feasible values of
$t$. Here, $p(\textbf{Z} \mid T_{ij} = t, \textbf{D}_{-ij}, \mathcal{I})$ is the likelihood of generating the
particular set of synthetic data given that $t$ is in the confidential
data and whatever else is known by the intruder.  The
$p(T_{ij} = t \mid \textbf{D}_{-ij}, \mathcal{I})$ can be considered the intruder's prior
distribution on $T_{ij}$ based on $(\textbf{D}_{-ij}, \mathcal{I})$.

As described in \citet{hureiterwang14}, intruders can use $p(T_{ij} = t \mid \textbf{Z}, \textbf{D}_{-ij}, \mathcal{I})$ to take guesses at
the true value $t_{ij}$.  For example, the intruder can find the
$t$ that offers the largest probability, and use that as a guess of
 $t_{ij}$.  Similarly, agencies can use $p(T_{ij} = t \mid \textbf{Z},
\textbf{D}_{-ij}, \mathcal{I})$ in disclosure risk evaluations. For example, for each
$t_{ij} \in \textbf{D}$, they can rank each $t$ by its associated value of
$p(T_{ij} = t \mid \textbf{Z}, \textbf{D}_{-ij}, \mathcal{I})$, and evaluate the rank at the truth,
$t = t_{ij}$.  When the rank of $t_{ij}$ is high (close to 1, which we
define to be the rank associated with the highest probability), the agency may deem
that record to be at risk under the strong intruder knowledge
scenario.  When the rank of $t_{ij}$ is low (far from 1), the agency may  deem
the risks for that record to be acceptable.

When $\mathcal{U}$ is very large, computing the normalizing constant in
\eqref{fullprob} is impractical. To facilitate
computation, we follow \citet{hureiterwang14} and consider as feasible candidates only
those $t$ that differ from $t_{ij}$ in one variable, along with
 $t_{ij}$ itself; we call this space $\mathcal{R}_{ij}$. Restricting to $\mathcal{R}_{ij}$ can be conceived as
mimicking a knowledgeable intruder who searches in spaces near $t_{ij}$.
As discussed by \citet{hureiterwang14}, restricting support to $\mathcal{R}_{ij}$ results in a conservative
ranking of the $t \in \mathcal{R}_{ij}$, in that ranks determined to be acceptably low
when using $\mathcal{R}_{ij}$ also are acceptably low when using $\mathcal{U}$.

For $T_i$, we use a similar approach to risk assessment. We compute
\begin{equation}
\rho_i^t = p(T_{i} = t \mid \textbf{Z}, \textbf{D}_{-i}, \mathcal{I}) \propto p(\textbf{Z} \mid T_{i}=t, \textbf{D}_{-i}, \mathcal{I})p(T_{i}=t
\mid \textbf{D}_{-i}, \mathcal{I}). \label{probpropAA}
\end{equation}
We consider only $t$ that differ from $t_i$ in either (i) one household-level variable for the entire household or (ii) one
individual-level variable for one household member, along with $t_i$ itself; we call
this space $\mathcal{R}_i$.

\section{Computational methods for risk assessment with the NDPMPM model}

We describe the computational methods for computing \eqref{probpropAA} in detail. Methods for computing
\eqref{propprob} are similar.

For any proposed $t$, let $\textbf{D}_i^t= (T_i=t, \textbf{D}_{-i})$ be the plausible confidential dataset when $T_i=t$.
Because each $\textbf{Z}^{(l)}$ is generated independently, we have
\begin{equation}
P(\textbf{Z} \mid \textbf{D}_i^t, \mathcal{I}) = \prod_{l=1}^L P(\textbf{Z}^{(l)} \mid \textbf{D}_i^t, \mathcal{I}). \label{gall}\end{equation}
Hence, we need to compute each $P(\textbf{Z}^{(l)} \mid \textbf{D}_i^t, \mathcal{I})$.

Let $\Theta=\{\mathbf{\pi}, \mathbf{\omega}, \mathbf{\lambda}, \mathbf{\phi} \}$ denote parameters from a NDPMPM
model. We can write $P(\textbf{Z}^{(l)} \mid \textbf{D}_i^t, \mathcal{I})$ as
\begin{equation}
P(\textbf{Z}^{(l)} \mid \textbf{D}_i^t, \mathcal{I}) = \int p(\textbf{Z}^{(l)} \mid \textbf{D}_{i}^t ,\mathcal{I}, \Theta) p(\Theta \mid
\textbf{D}_{i}^t, \mathcal{I})d\Theta. \label{Zeqn}
\end{equation}
To compute \eqref{Zeqn}, we could sample many values of $\Theta$ that could have generated $\textbf{Z}^{(l)}$; that is,
we could sample $\Theta^{(r)}$ for $r=1, \dots, R$. For each
$\Theta^{(r)}$, we compute the probability of generating the
released $\textbf{Z}^{(l)}$.  We then average these
probabilities over the $R$ draws of $\Theta$.

Conceptually, to draw $\Theta$ replicates, we could re-estimate the NDPMPM model for
each $\textbf{D}_i^t$.  This quickly becomes computationally prohibitive.
Instead, we suggest
using the sampled values of $\Theta$ from $p(\Theta \mid \textbf{D})$ as
proposals for an importance sampling algorithm.
To set notation, suppose we seek to estimate the expectation of
some function $g(\Theta)$, where $\Theta$ has density $f(\Theta)$. Further
suppose that we have available a sample
$(\Theta^{(1)},\dots,\Theta^{(R)})$ from a convenient distribution
$f^*(\Theta)$ that slightly differs from $f(\Theta)$. We can estimate
$E_f(g(\Theta))$ using
\begin{equation}
E_f(g(\Theta)) \approx \sum_{r=1}^R
g(\Theta^{(r)})\frac{f(\Theta^{(r)})/f^*(\Theta^{(r)})}{\sum_{r=1}^R
  f(\Theta^{(r)})/f^*(\Theta^{(r)})}. \label{impsamp}
\end{equation}

Let $t_i^{*(l)}$ be the $i$th
household's values of all variables, including household-level and individual-level variables, in synthetic dataset $\textbf{Z}^{(l)}$, where $i=1, \dots,
n$ and $l=1, \dots, L$.  For each $\textbf{Z}^{(l)}$ and any proposed $t$, we define the $g(\Theta)$ in \eqref{impsamp}
to equal $cP(\textbf{Z}^{(l)} \mid \textbf{D}_i^t, \mathcal{I})$. We approximate the expectation of each $g(\Theta)$ with respect to
$f(\Theta) = f(\Theta \mid \textbf{D}_i^t,\mathcal{I})$. In doing so, for any sampled $\Theta^{(r)}$
we use
\begin{equation}
g(\Theta^{(r)}) = P(\textbf{Z}^{(l)} \mid \textbf{D}_i^t, \mathcal{I}, \Theta^{(r)}) = \prod_{i=1}^{n}\left(\sum_{g=1}^F \pi_g^{(r)}\{ \prod_{k=p+1}^{p+q}\lambda_{gt_{ik}^{*(l)}}^{(k)(r)}(\prod_{j=1}^{n_i}\sum_{m=1}^S \omega_{gm}^{(r)}\prod_{k=1}^p\phi_{gmt_{ijk}^{*(l)}}^{(k)(r)})\}\right).
\end{equation}

We set $f^*(\Theta) = f(\Theta \mid \textbf{D}, \mathcal{I})$, so that we can use $R$ draws of $\Theta$ from
its posterior distribution based on $\textbf{D}$. Let these $R$ draws be $(\Theta^{(1)},\dots,\Theta^{(R)})$.
 We note that one could use any $\textbf{D}_{i}^t$ to obtain the $R$ draws, so that intruders can use similar importance sampling computations.
As evident in (1), (2), (3) and (4) in the main text, 
the only differences in the kernels of
$f(\Theta)$ and $f^*(\Theta)$ include (i) the components of the
likelihood associated with record $i$ and (ii) the normalizing constant for each density. Let $\textbf{t}=\{(c_{p+1},\dots,c_{p+q}),(c_{j1},\dots,c_{jp}), j=1,\dots,n_i\}$, where each $c_{k}\in(1,\dots,d_{k})$, be a guess at $T_i$, for household-level and individual-level variables respectively. After
computing the normalized ratio in \eqref{impsamp} and canceling common terms from the numerator and denominator, we are left with
$P(\textbf{Z}^{(l)} \mid \textbf{D}_i^t, \mathcal{I}) = \sum_{r=1}^R p_r q_r$ where

\begin{eqnarray}
 \label{pqhforHH-p} p_r &=& \prod_{i=1}^{n}\left(\sum_{g=1}^F \pi_g^{(r)}\{ \prod_{k=p+1}^{p+q}\lambda_{gt_{ik}^{*(l)}}^{(k)(r)}(\prod_{j=1}^{n_i}\sum_{m=1}^S \omega_{gm}^{(r)}\prod_{k=1}^p\phi_{gmt_{ijk}^{*(l)}}^{(k)(r)})\}\right)\\
  \label{pqhforHH-q} q_r &=& \frac{\frac{\sum_{g=1}^F \pi_g^{(r)}\{ \prod_{k=p+1}^{p+q}\lambda_{gc_{k}}^{(k)(r)}(\prod_{j=1}^{n_i}\sum_{m=1}^S \omega_{gm}^{(r)}\prod_{k=1}^p\phi_{gme_{jk}}^{(k)(r)})\}}{\sum_{g=1}^F \pi_g^{(r)}\{ \prod_{k=p+1}^{p+q}\lambda_{gt_{ik}}^{(k)(r)}(\prod_{j=1}^{n_i}\sum_{m=1}^S \omega_{gm}^{(h)}\prod_{k=1}^p\phi_{gmt_{ijk}}^{(k)(r)})\}}}
{\sum_{u=1}^R\left(\frac{\sum_{g=1}^F \pi_g^{(u)}\{ \prod_{k=p+1}^{p+q}\lambda_{gc_{k}}^{(k)(u)}(\prod_{j=1}^{n_i}\sum_{m=1}^S \omega_{gm}^{(u)}\prod_{k=1}^p\phi_{gme_{jk}}^{(k)(u)})\}}{\sum_{g=1}^F \pi_g^{(u)}\{ \prod_{k=p+1}^{p+q}\lambda_{gt_{ik}}^{(k)(u)}(\prod_{j=1}^{n_i}\sum_{m=1}^S \omega_{gm}^{(u)}\prod_{k=1}^p\phi_{gmt_{ijk}}^{(k)(u)})\}}\right)}.
\end{eqnarray}
We repeat this computation for each $\textbf{Z}^{(l)}$, plugging the $L$
results into \eqref{gall}.

 Finally, to approximate $\rho_i^t$, we compute
  \eqref{gall} for each $t \in \mathcal{R}_i$, multiplying each resulting value by its associated
  $P(T_i=t \mid \textbf{D}_{-i}, \mathcal{I})$. In what follows,  we presume an intruder with a uniform prior distribution over
the support $t \in \mathcal{R}_i$.  In this case, the prior probabilities cancel from the numerator and denominator of
\eqref{fullprob}, so that risk evaluations are based only on the likelihood function for $\textbf{Z}$.
We discuss evaluation of other prior distributions in the illustrative application.

For risk assessment for $T_{ij}$ in \eqref{propprob}, we use a similar importance sampling approximation, resulting in
\begin{eqnarray}
  \label{pqhforind-q} q_r &=& \frac{\frac{\sum_{g=1}^F \pi_g^{(r)}\{ \prod_{k=p+1}^{p+q}\lambda_{gc_{k}}^{(k)(r)}(\sum_{m=1}^S \omega_{gm}^{(r)}\prod_{k=1}^p\phi_{gme_{jk}}^{(k)(h)})\}}{\sum_{g=1}^F \pi_g^{(r)}\{ \prod_{k=p+1}^{p+q}\lambda_{gt_{ik}}^{(k)(h)}(\sum_{m=1}^S \omega_{gm}^{(r)}\prod_{k=1}^p\phi_{gmt_{ijk}}^{(k)(h)})\}}}
{\sum_{u=1}^R\left(\frac{\sum_{g=1}^F \pi_g^{(u)}\{ \prod_{k=p+1}^{p+q}\lambda_{gc_{k}}^{(k)(u)}(\sum_{m=1}^S \omega_{gm}^{(u)}\prod_{k=1}^p\phi_{gme_{jk}}^{(k)(u)})\}}{\sum_{g=1}^F \pi_g^{(u)}\{ \prod_{k=p+1}^{p+q}\lambda_{gt_{ik}}^{(k)(u)}(\sum_{m=1}^S \omega_{gm}^{(u)}\prod_{k=1}^p\phi_{gmt_{ijk}}^{(k)(u)})\}}\right)}.
\end{eqnarray}

\section{Disclosure risk assessments for synthesis without structural zeros}

To evaluate the  disclosure risks for individuals, we drop each individual record in $\textbf{D}$ one at a time.  For each individual $ij$, we compute
the resulting $\rho_{ij}^t$ for all $t$ in the reduced support $\mathcal{R}_{ij}$.
Here, each $\mathcal{R}_{ij}$ is the union of the true $t_{ij}$ plus the 39 other combinations of $t$ obtained by changing
one variable in $t_{ij}$ to any possible outcome.  For any two records $ij$ and $i'j'$ such that $t_{ij}
= t_{i'j'}$ in $\textbf{D}$,  $\rho_{ij}^t = \rho_{i'j'}^t$ for any possible $t$.
Thus, we need only compute the set of $\rho_{ij}^t$ for the 15280 combinations that appeared in $\textbf{D}$.  We use
a uniform prior distribution over all $t \in \mathcal{R}_{ij}$, for each record $ij$.

\begin{figure}[H]
\centering
\includegraphics[scale=0.28]{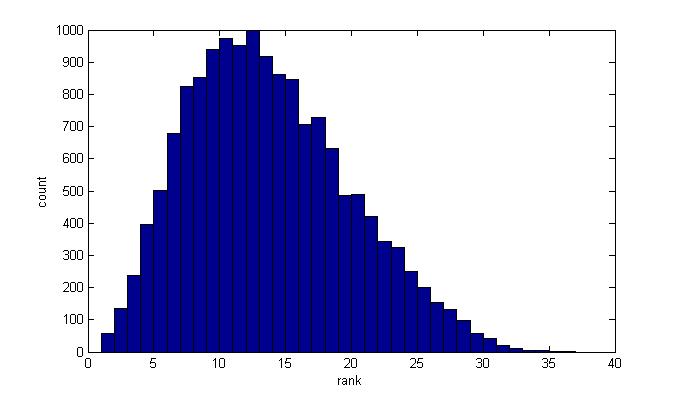}
\caption{Histogram of ranks of the probabilities associated with true $t_{ij}$.  Data have no structural zeros.}
\label{app1indirankcount}
\end{figure}

Figure \ref{app1indirankcount} displays the distribution of the rank of the true $t_{ij}$ for each of the 15280 combinations.
Here, a rank equal to 1 means the true $t_{ij}$ has the highest probability of being the unknown $T_{ij}$, whereas a rank of 40
means the true $t_{ij}$ has the lowest probability of being $T_{ij}$.  As evident in the figures, even armed with $\textbf{D}_{-ij}$
the intruder gives the top rank to the true $t_{ij}$ for only 11 combinations. The intruder gives $t_{ij}$ a ranking
in the top three for only 194 combinations. We note that, even though 12964 combinations were unique in $\textbf{D}$,
the NDPMPM synthesizer involves enough smoothing that we do not recover the true $t_{ij}$ in the overwhelming majority of cases.

\begin{figure}[H]
\centering
\includegraphics[scale=0.28]{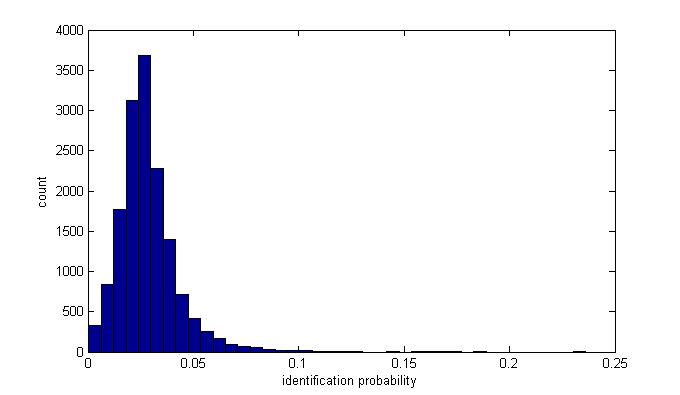}
\caption{Histogram of re-normalized probabilities associated with the true $t_{ij}$.  Data have no structural zeros.}
\label{app1indiidenprob}
\end{figure}

Figure \ref{app1indiidenprob} displays a histogram of the corresponding probabilities associated with the true $t_{ij}$ in each of the
15280 combinations. The largest probability is 0.2360. Only 1 probability exceeds 0.2, and 40 probabilities exceed 0.1.
The majority of probabilities are in the 0.03 range. As we assumed a uniform prior distribution over the 40 possibilities in $R_{ij}$,
the ratio of the posterior to prior probability is typically around one.  Only a handful of combinations have ratios exceeding two.
Thus, compared to random guesses over a close neighborhood of the true values, $\textbf{Z}$ typically does not provide much additional
information about $t_{ij}$.
We also look at the disclosure risks for households.  To do so, we drop each household record in $\textbf{D}$ one at a time.
 For households of size 2, the reduced support $\mathcal{R}_i$ comprises the true $t_i$ plus the 56 other combinations of $t$
obtained by changing $t_i$ in one variable. For the household-level variables, we change the entire variable for all members of the
household.  For the individual-level variable, we change one variable for each individual as before.
We need only compute $\rho_i^t$ for each of the 5375 combinations of households of size 2 that appear in $\textbf{D}$.
We use a uniform prior distribution over all $t \in \mathcal{R}_i$.

\begin{multicols}{2}

\end{multicols}

\begin{multicols}{2}
\begin{figure}[H]
\centering
\includegraphics[scale=0.28]{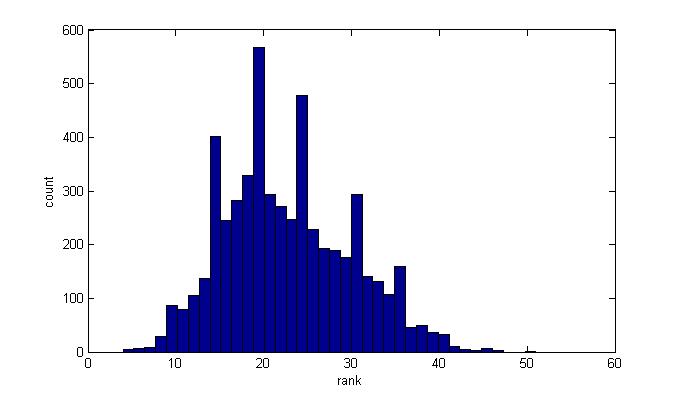}
\caption{Histogram of ranks of the probabilities associated with true $t_i$, for households of size 2.  Data have no structural zeros.}
\label{app1HHsize2rankcount}
\end{figure}

\begin{figure}[H]
\centering
\includegraphics[scale=0.28]{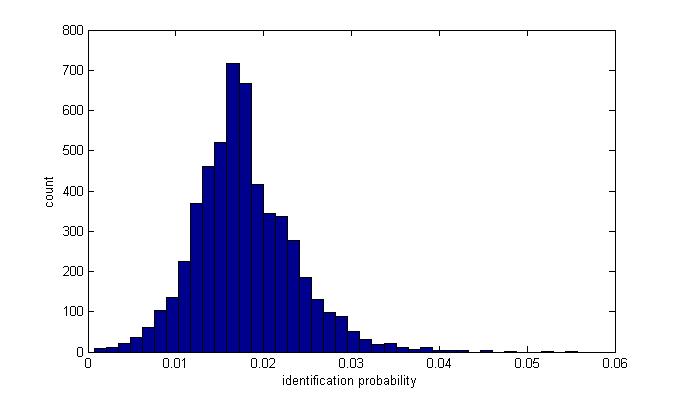}
\caption{Histogram of re-normalized probabilities associated with the true $t_i$, for households of size 2.  Data have no structural zeros.}
\label{app1HHsize2idenprob}
\end{figure}
\end{multicols}

Figure \ref{app1HHsize2rankcount} displays the distribution of the rank of the true $t_i$ for each of the 5375 combinations.
Once again, even armed with $\textbf{D}_{-i}$, the intruder never gives the top rank to the true $t_i$.  the
intruder gives the true $t_i$ a ranking in the top three for only seven household combinations. We note that 5331
household combinations of size 2 were unique in $\textbf{D}$.

Figure \ref{app1HHsize2idenprob} displays a histogram of the corresponding probabilities associated with the true $t_i$
in each of the 5375 combinations of households of size 2. The majority of probabilities are in the 0.02 range. As we
assumed a uniform prior distribution over the 57 possibilities in the support, the ratio of the posterior
to prior probability is typically around one. Thus, as with individuals, compared to random guesses over a
 close neighborhood of the true values,  $\textbf{Z}$ typically does not provide much additional information about $t_i$.
The largest probability is 0.0557.

For households of size 3, the reduced support $\mathcal{R}_i$ comprises the true $t_i$ plus 81 other combinations of $t$
obtained by changing one variable at a time, as done for households of size 3. We need only
compute $\rho_{i}^t$ for each of the 1375 combinations that appear in $\textbf{D}$. We use a uniform prior distribution over all $t \in \mathcal{R}_i$.

\begin{figure}[H]
\centering
\includegraphics[scale=0.28]{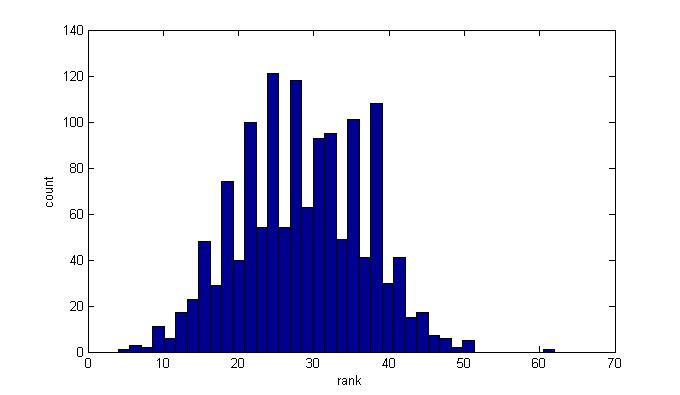}
\caption{Histogram of ranks of the probabilities associated with true $t_i$, for households of size 3.  Data have no structural zeros.}
\label{app1HHsize3rankcount}
\end{figure}

Figure \ref{app1HHsize3rankcount} displays the distribution of the rank of the true $t_i$ for each of the 1375 combinations. Even armed with $\textbf{D}_{-i}$, the intruder gives $t_i$ a ranking in the top three for no combinations. We note that all these 1375 combinations were unique in $\textbf{D}$, yet evidently the nested Dirichlet process synthesizer involves enough smoothing that we do not recover the true $t_i$ in the overwhelming majority of cases.

\begin{multicols}{2}

\begin{figure}[H]
\centering
\includegraphics[scale=0.28]{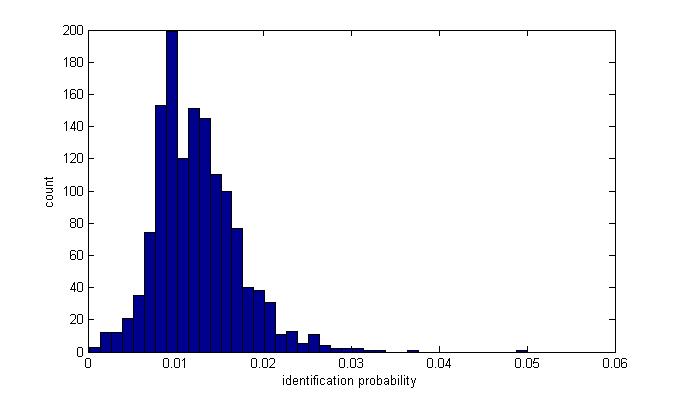}
\caption{Histogram of re-normalized probabilities associated with the true $t_i$, for households of size 3.  Data have no structural zeros.}
\label{app1HHsize3idenprob}
\end{figure}
\end{multicols}

Figure \ref{app1HHsize3idenprob} displays a histogram of the corresponding probabilities associated with the true $t_i$ in each of the 1375 combinations. The majority of probabilities are in the 0.010 range. As we assumed a uniform prior distribution over the 82 possibilities in the support, the ratio of the posterior to prior probability is typically one or less. Thus, compared to random guesses over a reasonably close neighborhood of the true values,  $\textbf{Z}$ typically does not provide much additional information about $t_i$. The largest probability is 0.0500.

For households of size 4, the reduced support $\mathcal{R}_i$ comprises the true $t_i$ plus 106 other combinations of $t$
obtained by changing one variable at a time, as with the other sizes. We do computations for each of the 478 combinations
that appear in $\textbf{D}$. We use a uniform prior distribution over all $t \in \mathcal{R}_i$.

\begin{multicols}{2}
\begin{figure}[H]
\centering
\includegraphics[scale=0.28]{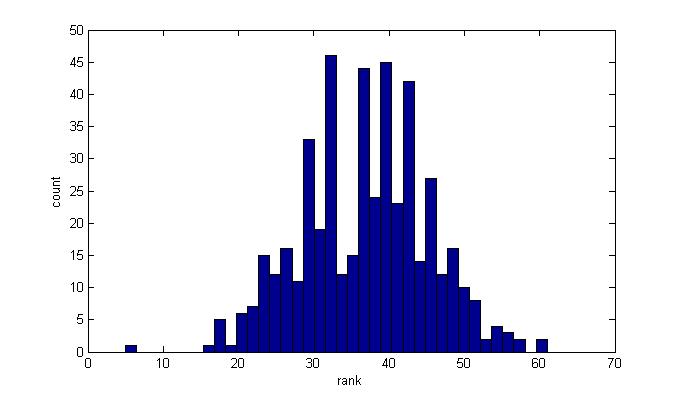}
\caption{Histogram of ranks of the probabilities associated with true $t_i$, for households of size 4.  Data have no structural zeros.}
\label{app1HHsize4rankcount}
\end{figure}

\begin{figure}[H]
\centering
\includegraphics[scale=0.28]{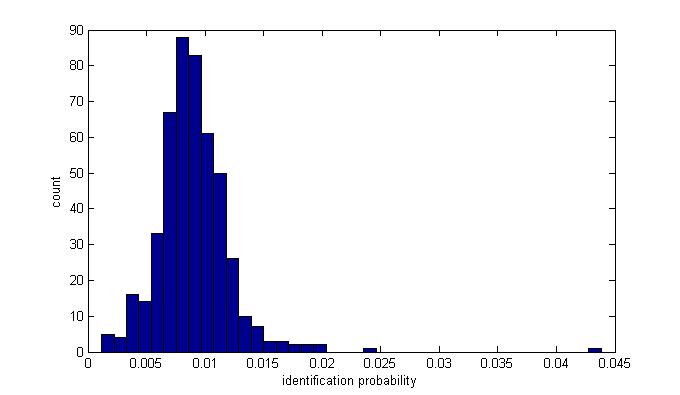}
\caption{Histogram of re-normalized probabilities associated with the true $t_i$, for households of size 4.  Data have no structural zeros.}
\label{app1HHsize4idenprob}
\end{figure}
\end{multicols}

Figure \ref{app1HHsize4rankcount} displays the distribution of the rank of the true $t_i$ for each of the 478 combinations.
The intruder gives the true $t_i$ a ranking in the top three for no combinations. All these 478 combinations were unique in $\textbf{D}$.
Figure \ref{app1HHsize4idenprob} displays a histogram of the corresponding probabilities associated with the true $t_i$ in each
of the 478 combinations. The majority of probabilities are in the 0.01 range. As we assumed a uniform prior distribution over the
107 possibilities in the support, the ratio of the
posterior to prior probability is typically one or less. Once again, $\textbf{Z}$ typically does not provide much additional information about $t_i$. The largest probability is 0.0438.

For households of size 5, the reduced support $\mathcal{R}_i$ comprises the true $t_i$ plus 131 other combinations of $t$
obtained by changing one variable at a time, as with the other sizes. We do computations for each of the 123 combinations
that appear in $\textbf{D}$. We use a uniform prior distribution over all $t \in \mathcal{R}_i$.

\begin{figure}[H]
\centering
\includegraphics[scale=0.28]{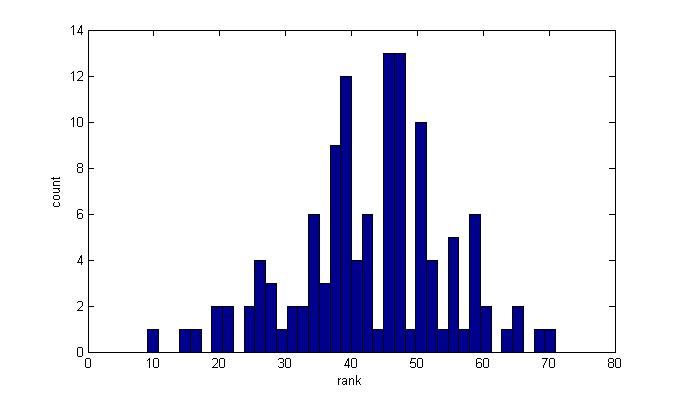}
\caption{Histogram of ranks of the probabilities associated with true $t_i$, for households of size 5.  Data have no structural zeros.}
\label{app1HHsize5rankcount}
\end{figure}

\begin{figure}[H]
\centering
\includegraphics[scale=0.28]{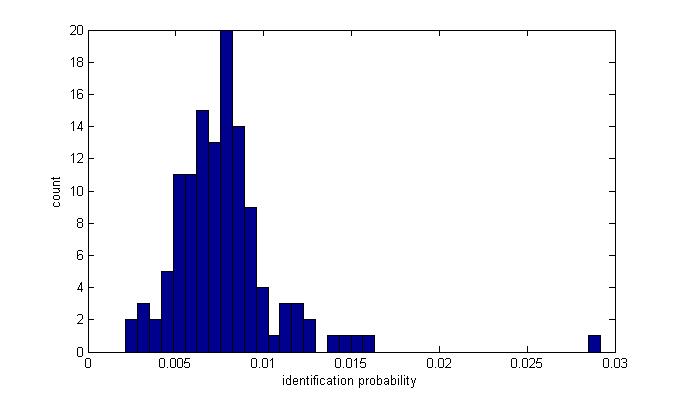}
\caption{Histogram of re-normalized probabilities associated with the true $t_i$, for households of size 5.  Data have no structural zeros.}
\label{app1HHsize5idenprob}
\end{figure}

Figure \ref{app1HHsize5rankcount} displays the distribution of the rank of the true $t_i$ for each of the 123 combinations.
The intruder gives the true $t_i$ a ranking in the top three for no combinations. All these 123 combinations were unique in $\textbf{D}$.
Figure \ref{app1HHsize5idenprob} displays a histogram of the corresponding probabilities associated with the true $t_i$ in each
of the 123 combinations. The majority of probabilities are in the 0.008 range. As we assumed a uniform prior distribution over the
132 possibilities in the support, the ratio of the
posterior to prior probability is typically around one. Once again, $\textbf{Z}$ typically does not provide much additional information about $t_i$. The largest probability is 0.0292.

\begin{multicols}{2}

\end{multicols}

For households of size 6, the reduced support $\mathcal{R}_i$ comprises the true $t_i$ plus 156 other combinations of $t$
obtained by changing one variable at a time, as with the other sizes. We do computations for each of the 52 combinations
that appear in $\textbf{D}$. We use a uniform prior distribution over all $t \in \mathcal{R}_i$.

\begin{multicols}{2}
\begin{figure}[H]
\centering
\includegraphics[scale=0.28]{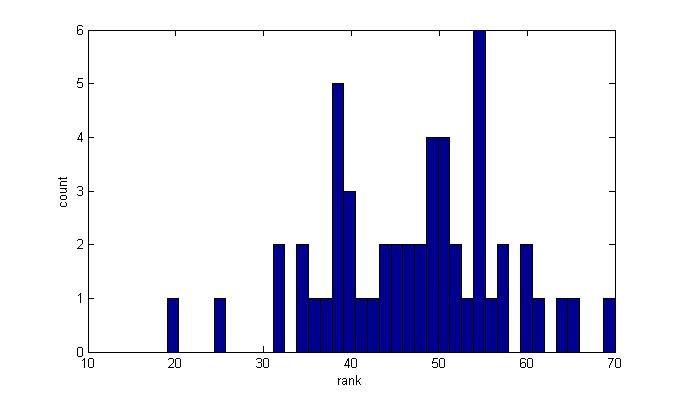}
\caption{Histogram of ranks of the probabilities associated with true $t_i$, for households of size 6.  Data have no structural zeros.}
\label{app1HHsize6rankcount}
\end{figure}

\begin{figure}[H]
\centering
\includegraphics[scale=0.28]{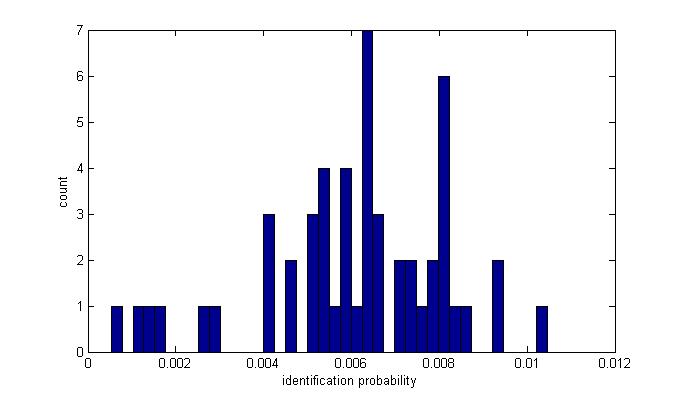}
\caption{Histogram of re-normalized probabilities associated with the true $t_i$, for households of size 6.  Data have no structural zeros.}
\label{app1HHsize6idenprob}
\end{figure}
\end{multicols}

Figure \ref{app1HHsize6rankcount} displays the distribution of the rank of the true $t_i$ for each of the 52 combinations.
The intruder gives the true $t_i$ a ranking in the top three for no combinations. All these 52 combinations were unique in $\textbf{D}$.
Figure \ref{app1HHsize6idenprob} displays a histogram of the corresponding probabilities associated with the true $t_i$ in each
of the 52 combinations. The majority of probabilities are in the 0.007 range. As we assumed a uniform prior distribution over the
157 possibilities in the support, the ratio of the
posterior to prior probability is typically around one. Once again, $\textbf{Z}$ typically does not provide much additional information about $t_i$. The largest probability is 0.0105.

\begin{multicols}{2}
\begin{figure}[H]
\centering
\includegraphics[scale=0.28]{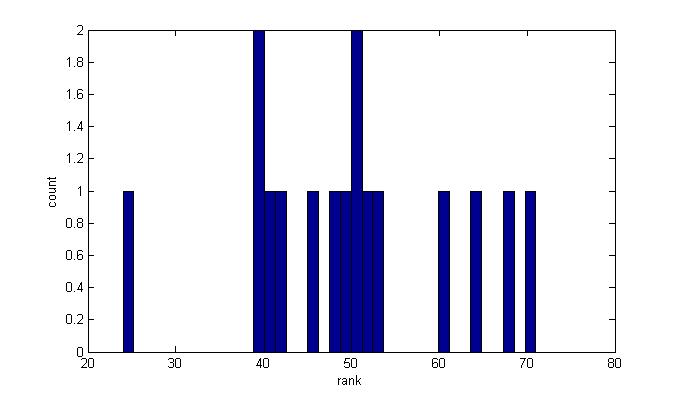}
\caption{Histogram of ranks of the probabilities associated with true $t_i$, for households of size 7.  Data have no structural zeros.}
\label{app1HHsize7rankcount}
\end{figure}

\begin{figure}[H]
\centering
\includegraphics[scale=0.28]{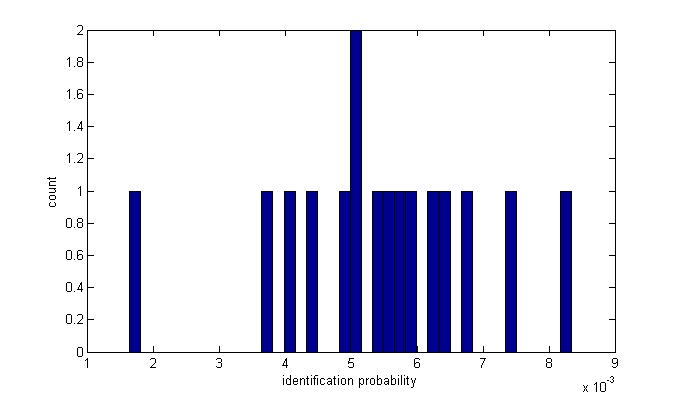}
\caption{Histogram of re-normalized probabilities associated with the true $t_i$, for households of size 7.  Data have no structural zeros.}
\label{app1HHsize7idenprob}
\end{figure}
\end{multicols}

For households of size 7, the reduced support $\mathcal{R}_i$ comprises the true $t_i$ plus 181 other combinations of $t$
obtained by changing one variable at a time, as with the other sizes. We do computations for each of the 16 combinations
that appear in $\textbf{D}$. We use a uniform prior distribution over all $t \in \mathcal{R}_i$.

Figure \ref{app1HHsize7rankcount} displays the distribution of the rank of the true $t_i$ for each of the 16 combinations.
The intruder gives the true $t_i$ a ranking in the top three for no combinations. All these 16 combinations were unique in $\textbf{D}$.
Figure \ref{app1HHsize7idenprob} displays a histogram of the corresponding probabilities associated with the true $t_i$ in each
of the 16 combinations. The majority of probabilities are in the 0.005 range. As we assumed a uniform prior distribution over the
182 possibilities in the support, the ratio of the
posterior to prior probability is typically around one. Once again, $\textbf{Z}$ typically does not provide much additional information about $t_i$. The largest probability is 0.0083.

For households of size 8, the reduced support $\mathcal{R}_i$ comprises the true $t_i$ plus 206 other combinations of $t$
obtained by changing one variable at a time, as with the other sizes. We do computations for each of the 5 combinations
that appear in $\textbf{D}$. We use a uniform prior distribution over all $t \in \mathcal{R}_i$.

The ranks of the true $t_i$ for each of the 5 combinations are $\{52, 39, 84, 57, 67\}$. The intruder gives the true $t_i$ a ranking in the top three for no combinations. We note that all 5
household combinations of size 8 were unique in $\textbf{D}$.
The corresponding probabilities associated with the true $t_i$ in each
of the 4 combinations are $\{0.0057, 0.0049, 0.0043, 0.0075, $ $0.0041\}$. As we assumed a uniform prior distribution over the
207 possibilities in the support, the ratio of the
posterior to prior probability is typically around one. Once again, $\textbf{Z}$ typically does not provide much additional information about $t_i$. The largest probability is 0.0075.

For households of size 9, the reduced support $\mathcal{R}_i$ comprises the true $t_i$ plus 231 other combinations of $t$
obtained by changing one variable at a time, as with the other sizes. We do computations for each of the 2 combinations
that appear in $\textbf{D}$. We use a uniform prior distribution over all $t \in \mathcal{R}_i$.

The ranks of the true $t_i$ for each of the 2 combinations are $\{57, 66\}$. We note that both 2
household combinations of size 9 were unique in $\textbf{D}$.
The corresponding probabilities associated with the true $t_i$ in each
of the 2 combinations are $\{0.0029, 0.0017\}$. As we assumed a uniform prior distribution over the
232 possibilities in the support, the ratio of the
posterior to prior probability is less than one. Once again, $\textbf{Z}$ typically does not provide much additional information about $t_i$.

\section{Disclosure risk assessments for structural zeros example}

We now turn to illustrating the assessment of disclosure risks for the synthesis with structural zeros, described in Section 4.2 of the main text.
For individual disclosure risks, for each individual $ij$ we compute the $\rho_{ij}^t$ for all $t$ in $\mathcal{R}_{ij}$
defined as the union of the true $t_{ij}$ plus the 24 other combinations of $t$ obtained by changing
one variable at a time, keeping the relationship variable fixed as a computational convenience.
We compute $\rho_{ij}^t$ for each of the 2517 combinations that appear in $\textbf{D}$.  We use a uniform prior distribution over all $t \in \mathcal{R}_{ij}$.
\begin{multicols}{2}
\begin{figure}[H]
\centering
\includegraphics[scale=0.28]{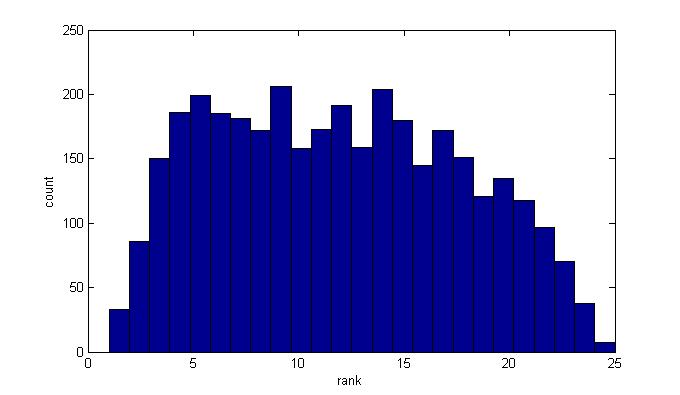}
\caption{Histogram of ranks of the probabilities associated with true $t_{ij}$.  Data have structural zeros.}
\label{app2indirankcount}
\end{figure}

\begin{figure}[H]
\centering
\includegraphics[scale=0.28]{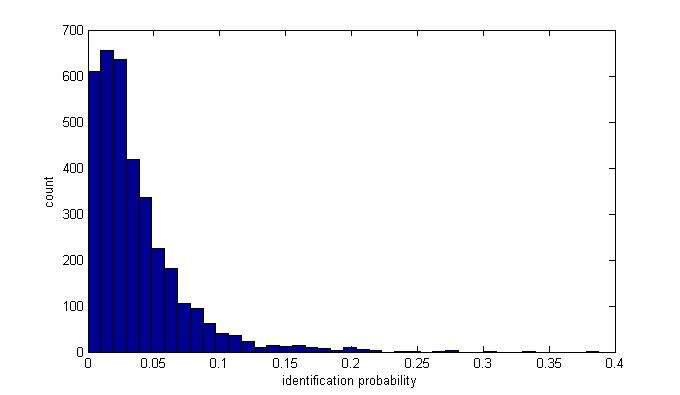}
\caption{Histogram of re-normalized probabilities associated with the true $t_{ij}$, all individuals. Data have structural zeros.}
\label{app2indiidenprob}
\end{figure}
\end{multicols}

Figure \ref{app2indirankcount} displays the distribution of the rank of the true $t_{ij}$ for each of the 3517 combinations. Even armed
with $\textbf{D}_{-ij}$, the intruder gives the top rank to the true $t_{ij}$ for only 33 combinations.  The intruder gives the true $t_{ij}$ a ranking
in the top three for 269 combinations.  We note that 1204 combinations were unique in $\textbf{D}$.

Figure \ref{app2indiidenprob} displays a histogram of the corresponding probabilities associated with the true $t_i$ in each of the
3517 combinations. The majority of probabilities are in the 0.03 range. As we assumed a uniform prior distribution over the 25 possibilities in the support,
the ratio of the posterior to prior probability is typically only slightly above one. Thus, compared to random guesses over a close neighborhood
of the true values,  $\textbf{Z}$ typically does not provide much additional information about $t_{ij}$.
The largest probability is 0.3878, and only 4 probabilities exceed 0.3, 27 probabilities exceed 0.2, and 183 probabilities exceed 0.1.

We also look at the disclosure risks for households. For households of size 2, the reduced support $\mathcal{R}_i$ consists of the true $t_i$ plus 31 other combinations of $t$ obtained by changing $t_i$ in one variable. We need only do computations for each of the 4070 combinations that appeared in $\textbf{D}$. We use a uniform prior distribution over all $t \in \mathcal{R}_i$.

\begin{figure}[H]
\centering
\includegraphics[scale=0.28]{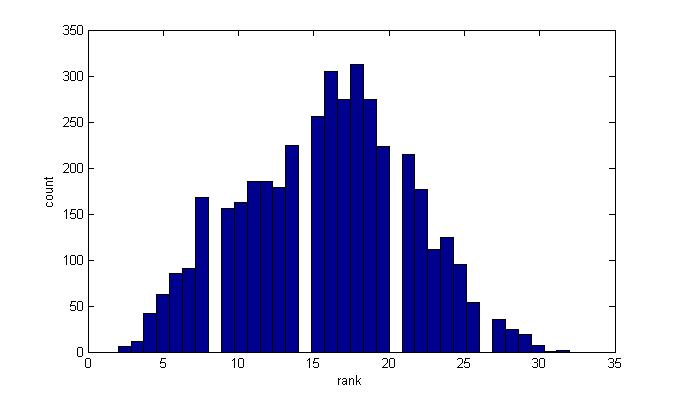}
\caption{Histogram of ranks of the probabilities associated with true $t_i$, for households of size 2.  Data have structural zeros.}
\label{app2HHsize2rankcount}
\end{figure}

Figure \ref{app2HHsize2rankcount} displays the distribution of the rank of the true $t_i$ for each of the 4070 combinations. Even armed with $\textbf{D}_{-i}$, the intruder gives the top rank to the true $t_i$ for no household combination, and gives $t_i$ a ranking
in the top three for only 18 combinations. We note that 3485 combinations were unique in $\textbf{D}$.

\begin{figure}[H]
\centering
\includegraphics[scale=0.28]{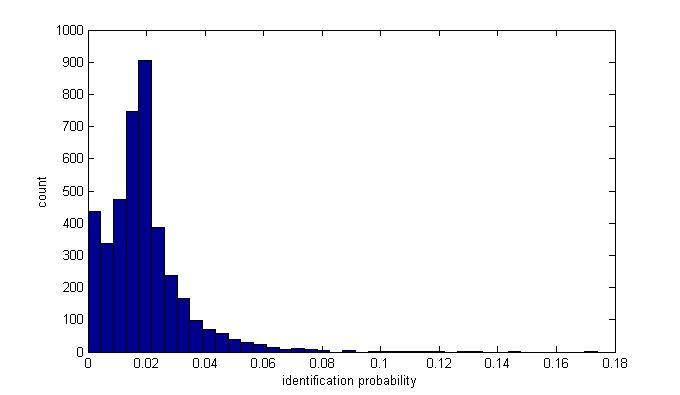}
\caption{Histogram of re-normalized probabilities associated with the true $t_i$, for households of size 2.  Data have structural zeros.}
\label{app2HHsize2idenprob}
\end{figure}

Figure \ref{app2HHsize2idenprob} displays a histogram of the corresponding probabilities associated with the true $t_i$ in each of the
4070 combinations. The majority of probabilities are in the 0.025 range. As we assumed a uniform prior distribution over the 32 possibilities in the support,
the ratio of the posterior to prior probability is typically one or less. Thus, compared to random guesses over a reasonably close neighborhood
of the true values,  $\textbf{Z}$ typically does not provide much additional information about $t_i$. The largest probability is 0.1740, and only 15 probabilities exceed 0.1.

\begin{multicols}{2}

\end{multicols}
For households of size 3, the reduced support $\mathcal{R}_i$ consists of the true $t_i$ plus 46 other combinations of $t$. We need only do computations for each of the 2492 combinations that appeared in $\textbf{D}$. We use a uniform prior distribution over all $t \in \mathcal{R}_i$.

\begin{multicols}{2}
\begin{figure}[H]
\centering
\includegraphics[scale=0.28]{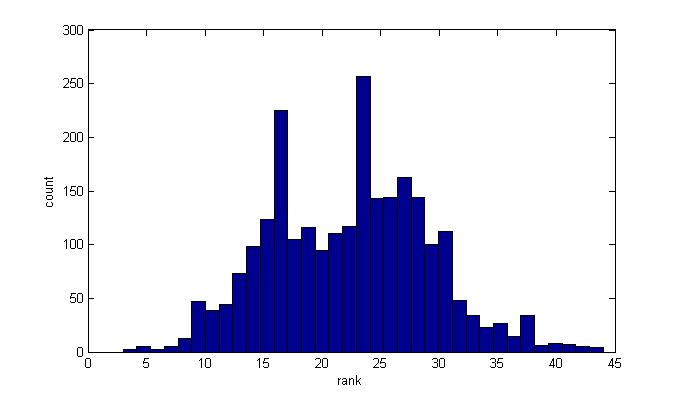}
\caption{Histogram of ranks of the probabilities associated with true $t_i$, for households of size 3.  Data have structural zeros.}
\label{app2HHsize3rankcount}
\end{figure}

\begin{figure}[H]
\centering
\includegraphics[scale=0.28]{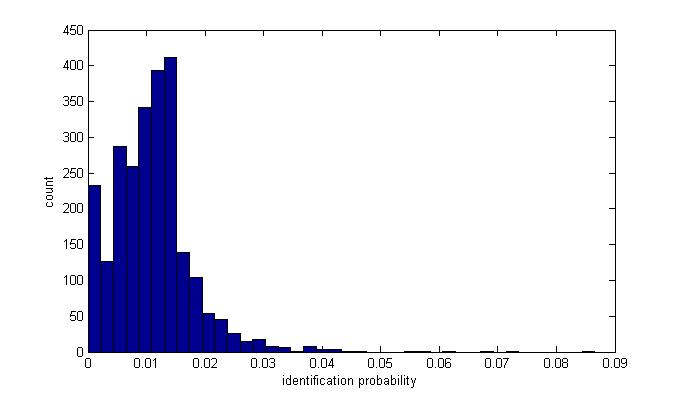}
\caption{Histogram of re-normalized probabilities associated with the true $t_i$, for households of size 3.  Data have structural zeros.}
\label{app2HHsize3idenprob}
\end{figure}
\end{multicols}

Figure \ref{app2HHsize3rankcount} displays the distribution of the rank of the true $t_i$ for each of the 2492 combinations. Even armed with $\textbf{D}_{-i}$, the intruder gives the top rank to the true $t_i$ for no combination and gives $t_i$ a ranking
in the top three for only 2 combinations. We note that 2480 combinations were unique in $\textbf{D}$.

Figure \ref{app2HHsize3idenprob} displays a histogram of the corresponding probabilities associated with the true $t_i$ in each of the
2492 combinations. The majority of probabilities are in the 0.01 range. As we assumed a uniform prior distribution over the 47 possibilities in the support,
the ratio of the posterior to prior probability is typically less than one. Thus, compared to random guesses over a reasonably close neighborhood
of the true values,  $\textbf{Z}$ typically does not provide much additional information about $t_i$. The largest probability is 0.0866.

\begin{multicols}{2}
\begin{figure}[H]
\centering
\includegraphics[scale=0.28]{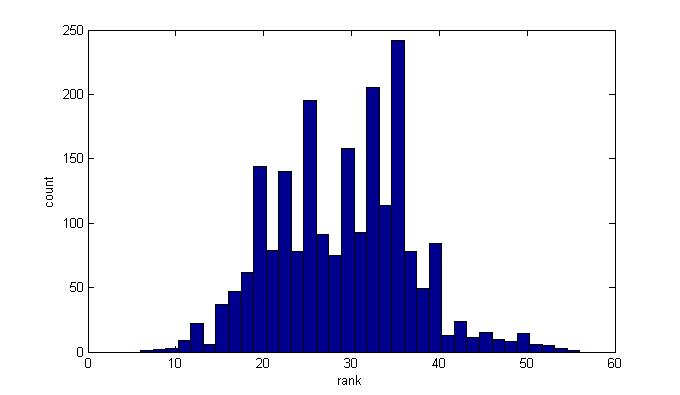}
\caption{Histogram of ranks of the probabilities associated with true $t_i$, for households of size 4.  Data have structural zeros.}
\label{app2HHsize4rankcount}
\end{figure}

\begin{figure}[H]
\centering
\includegraphics[scale=0.28]{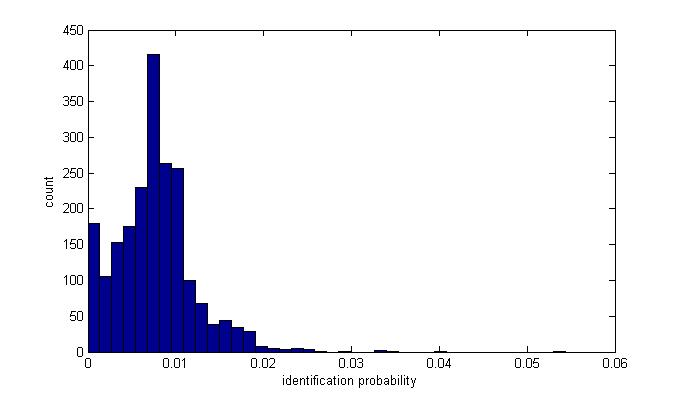}
\caption{Histogram of re-normalized probabilities associated with the true $t_i$, for households of size 4.  Data have structural zeros.}
\label{app2HHsize4idenprob}
\end{figure}
\end{multicols}

For households of size 4, the reduced support $\mathcal{R}_i$ consists of the true $t_i$ plus 61 other combinations of $t$. We need only do computations for each of the 2124 combinations that appeared in $\textbf{D}$. We use a uniform prior distribution over all $t \in \mathcal{R}_i$.

Figure \ref{app2HHsize4rankcount} displays the distribution of the rank of the true $t_i$ for each of the 2124 combinations. Even armed with $\textbf{D}_{-i}$, the intruder gives the top rank to the true $t_i$ for no combination and gives $t_i$ a ranking in the top three for no combinations. We note that 2122 combinations were unique in $\textbf{D}$.

Figure \ref{app2HHsize4idenprob} displays a histogram of the corresponding probabilities associated with the true $t_i$ in each of the
2124 combinations. The majority of probabilities are in the 0.01 range. As we assumed a uniform prior distribution over the 62 possibilities in the support,
the ratio of the posterior to prior probability is typically less than one. Thus, compared to random guesses over a reasonably close neighborhood
of the true values,  $\textbf{Z}$ typically does not provide much additional information about $t_i$. The largest probability is 0.0544.

\section{Synthetic data and original sample estimates versus population values}

\begin{figure}[H]
\centering
\includegraphics[scale=0.35]{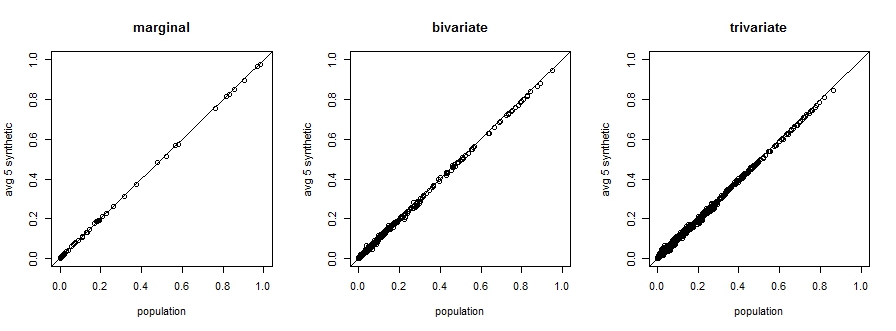}
\caption{Marginal, bivariate and trivariate probabilities computed in the population and synthetic
  datasets for the illustration without structural zeros.  Point
  estimates from the synthetic datasets and the population parameters are similar, suggesting that
  the NDPMPM estimates the population parameters well.}
\label{app1_1}
\end{figure}

\begin{figure}[H]
\centering
\includegraphics[scale=0.35]{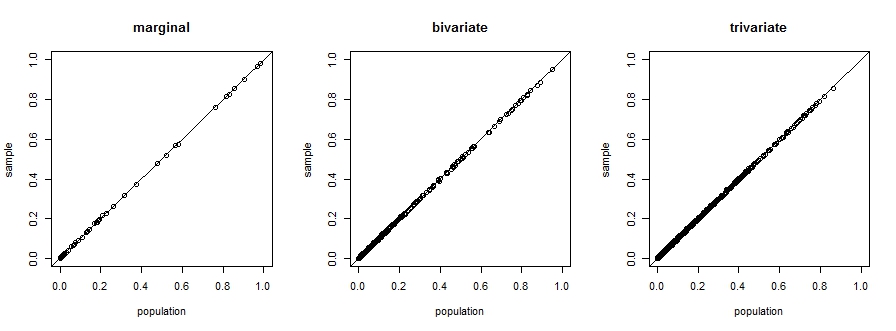}
\caption{Marginal, bivariate and trivariate probabilities computed in the population and the sample for the illustration without structural zeros.  Point estimates from the sample and the population parameters are similar.}
\label{app1_2}
\end{figure}

\begin{figure}[H]
\centering
\includegraphics[scale=0.35]{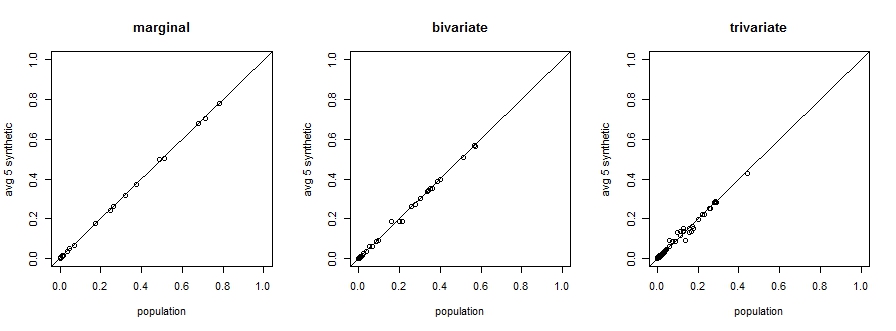}
\caption{Marginal, bivariate and trivariate distributions probabilities computed in the population and synthetic
  datasets in illustration with structural zeros.  Point
  estimates from the synthetic datasets and the population parameters are similar, suggesting that
  the NDPMPM estimates the population parameters well.}
\label{app2_1}
\end{figure}

\begin{figure}[H]
\centering
\includegraphics[scale=0.35]{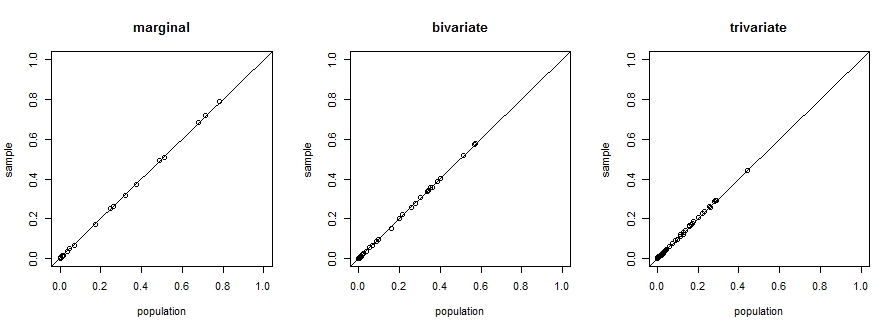}
\caption{Marginal, bivariate and trivariate distributions probabilities computed in the population and the sample in illustration with structural zeros.
 Point estimates from the sample and the population parameters are similar.}
\label{app2_2}
\end{figure}

In this section, we present plots of point estimates for the original samples versus the values in the constructed populations, and
for the synthetic data versus the values in the constructed populations.  Figure \ref{app1_1} and Figure \ref{app1_2} display plots for the no structural
 zeros simulation described in Section 4.1 of the main text. Figure \ref{app2_1} and Figure \ref{app2_2}
display plots for the structural zeros simulation described in the main text. In both simulation scenarios,
the synthetic data and the original sample point estimates are close to the population values.

\section{Uniform prior results in the no structural zeros simulation}

In the main text, we presented results based on using the empirical marginal frequencies as the shape parameters for the Dirichlet distributions
in the main text.  Here, we present results using uniform prior distributions for $\lambda$ and $\phi$ for the scenario with no structural zeros (Section 4.1 in the main text).

\begin{figure}[H]
\centering
\includegraphics[scale=0.35]{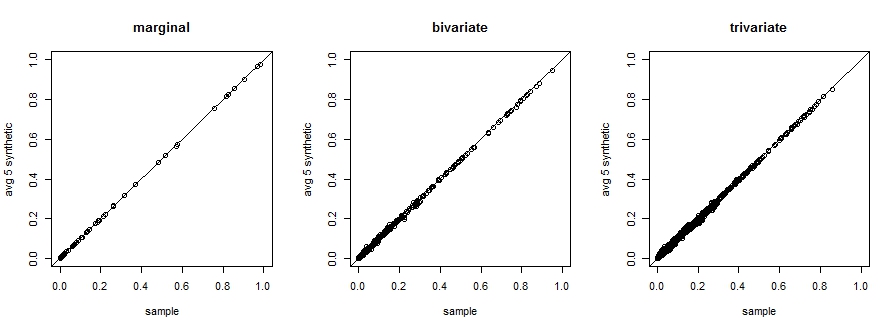}
\caption{Marginal, bivariate and trivariate probabilities computed in the sample and synthetic
  datasets for the illustration without structural zeros, using uniform prior for $\lambda$ and $\phi$.}
\label{app1_3}
\end{figure}

Figure \ref{app1_3} displays plots of point estimates with the uniform priors, which are very similar to the plots in Figure 1 in the main text
based on the empirical priors.  Table \ref{app1HHraceandtype_uniform} displays probabilities for  within-household relationships
for the model with the uniform prior distribution, along with the results based on the empirical prior distribution for comparison.
We find no meaningful differences between the two sets of results.

\begin{table}[H]
\centering
\begin{tabular}{lrrrr}
 &Q&Original &Uniform&Empirical\\ \hline
  All same race && &\\
$\,\,\,\,\,$ $n_i = 2$&.928 &(.923, .933)& (.840, .859)&(.847, .868)\\
$\,\,\,\,\,$  $n_i = 3$&.906 &(.889, .901)& (.809, .854)&(.803, .845)\\
$\,\,\,\,\,$ $n_i = 4$&.885&(.896, .908)&(.747, .831)&(.730, .817)\\
  All white and rent&.123 & (.115, .128)&(.110, .125)&(.110, .126)\\
  All white and have health coverage&.632 &(.622, .641)&(.579, .605)&(.582, .603)\\
  All married and working&.185 & (.177, .192)&(.163, .179)&(.171, .188)\\
  All have college degree&.091 &(.086, .097)& (.069, .080)&(.071, .082)\\
  All have health coverage&.807 &(.800, .815)& (.764, .784)&(.764, .782)\\
  All speak English&.974 &(.969, .976)&(.958, .966)&(.959, .967)\\
  Two workers in house &.291 &(.282, .300) & (.282, .304)&(.289, .309)\\  \hline
\end{tabular}
\caption{95\% confidence intervals in the original and synthetic data using a uniform prior and an empirical prior
  for selected probabilities that depend on within household
  relationships. Results for illustration without structural zeros.
  Intervals for probability that all family members are the same race
  are presented only for households of size two, three, and four
because of inadequate sample sizes for $n_i>4$.
The
quantity $Q$ is the value in the constructed population of 308769
households.}
\label{app1HHraceandtype_uniform}
\end{table}

\section{Results for larger number of components}

\begin{figure}[H]
\centering
\includegraphics[scale=0.35]{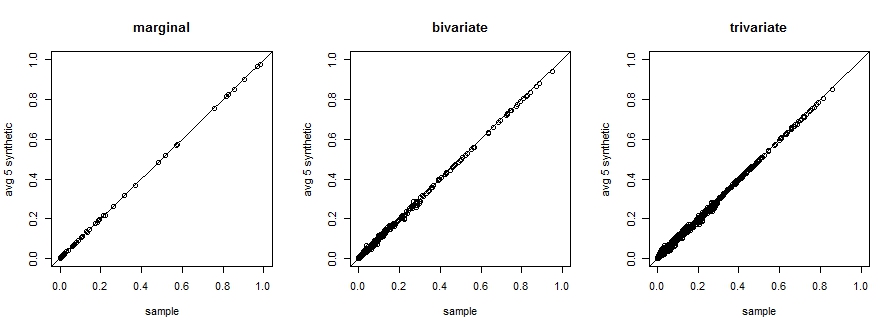}
\caption{Marginal, bivariate and trivariate probabilities computed in the sample and synthetic
  datasets for the illustration without structural zeros, with $(F, S) = (50, 50)$.}
\label{app1_4}
\end{figure}

In this section, we present results using $(F, S) = (50, 50)$ for the no structural zeros simulation, which results in many more classes than the results based on $(F, S) = (30, 10)$
that are presented in the main text. Figure \ref{app1_4} displays plots of point estimates with $(F, S) = (50, 50)$.
 These are very similar to the plots in Figure 1 in the main text. Table \ref{app1HHraceandtype_largeFS} displays probabilities
that depend on within-household relationships using these two sets of values of $(F, S)$. We find
no meaningful differences between these two sets of results.

\begin{table}[H]
\centering
\begin{tabular}{lrrrr}
 &Q&Original &$(50, 50)$ & $(30, 10)$\\ \hline
  All same race && &\\
$\,\,\,\,\,$ $n_i = 2$&.928 &(.923, .933)& (.835, .861)&(.847, .868)\\
$\,\,\,\,\,$  $n_i = 3$&.906 &(.889, .901)& (.820, .861)&(.803, .845)\\
$\,\,\,\,\,$ $n_i = 4$&.885&(.896, .908)&(.755, .845)&(.730, .817)\\
  All white and rent&.123 & (.115, .128)&(.110, .125)&(.110, .126)\\
  All white and have health coverage&.632 &(.622, .641)&(.583, .606)&(.582, .603)\\
  All married and working&.185 & (.177, .192)&(.168, .186)&(.171, .188)\\
  All have college degree&.091 &(.086, .097)& (.069, .080)&(.071, .082)\\
  All have health coverage&.807 &(.800, .815)& (.761, .784)&(.764, .782)\\
  All speak English&.974 &(.969, .976)&(.958, .967)&(.959, .967)\\
  Two workers in house &.291 &(.282, .300) & (.291, .313)&(.289, .309)\\  \hline
\end{tabular}
\caption{95\% confidence intervals in the original and synthetic data (using $(F, S) = (50, 50)$ and $(F, S) = (30, 10)$) for selected probabilities that depend on within household relationships. Results for illustration without structural zeros. Intervals for probability that all family members are the same race are presented only for households of size two, three, and four because of inadequate sample sizes for $n_i>4$. The quantity $Q$ is the value in the constructed population of 308769
households.}
\label{app1HHraceandtype_largeFS}
\end{table}

\bibliographystyle{ba}
\bibliography{disclosurebib}

\end{document}